\documentclass[10pt,aps,prd,twocolumn,floats,floatfix,showpacs,superscriptaddress,nofootinbib]{revtex4-1}

\usepackage[dvipsnames]{xcolor}
\usepackage{amsmath,amssymb}
\usepackage{bm}
\usepackage[colorlinks]{hyperref}
\usepackage{cleveref}
\usepackage[utf8]{inputenc}
\usepackage{graphicx}

\newcommand{\dif}{\mathrm{d}}

\begin{document}

\title{Black hole superradiant instability from ultralight spin-2 fields}
\author{Richard Brito}
	\email{richard.brito@roma1.infn.it}
	\affiliation{Dipartimento di Fisica, ``Sapienza" Università di Roma \& Sezione INFN Roma1, Piazzale Aldo Moro 5, 
00185, Roma, Italy}
\author{Sara Grillo}
	\affiliation{Dipartimento di Fisica G. Occhialini, Universit\`a degli Studi di Milano Bicocca, Piazza della Scienza 
3, 20126 Milano, Italy}
	\affiliation{Dipartimento di Fisica, ``Sapienza" Università di Roma \& Sezione INFN Roma1, Piazzale Aldo Moro 
5, 00185, Roma, Italy}
\author{Paolo Pani}
	\email{paolo.pani@uniroma1.it}
	\affiliation{Dipartimento di Fisica, ``Sapienza" Università di Roma \& Sezione INFN Roma1, Piazzale Aldo Moro 5, 00185, Roma, Italy}

\begin{abstract}
Ultralight bosonic fields are compelling dark-matter candidates and arise in a variety of beyond-Standard-Model 
scenarios. These fields can tap energy and angular momentum from spinning 
black holes through superradiant instabilities, during which a macroscopic bosonic condensate develops 
around the black hole. Striking features of this phenomenon include gaps in the spin-mass distribution of 
astrophysical black holes and a continuous gravitational-wave~(GW) signal
emitted by the condensate.
So far these processes have been studied in great detail for scalar 
fields and, more recently, for vector fields. Here we take an important step forward in the black-hole superradiance 
program by computing, analytically, the instability time scale, the direct
GW emission, and the stochastic background, in the case of massive tensor (i.e., spin-$2$) fields. 
Our analysis is valid for any black hole spin and for small boson masses. The instability of massive spin-$2$ fields
shares some properties with the scalar and vector cases, but its phenomenology is much richer, for example there exist 
multiple modes with comparable instability time scales, and the dominant
GW signal is hexadecapolar rather than quadrupolar.
Electromagnetic and GW observations of spinning black holes in the mass range $M\in(1,10^{10})M_\odot$ can constrain the mass of a putative spin-$2$ field in the range $10^{-22} \lesssim m_b\,{\rm c^2/eV} \lesssim 10^{-10}$. For $10^{-17}\lesssim m_b\,{\rm c^2/eV}\lesssim  10^{-15}$, the 
space mission LISA could detect the continuous GW signal for sources at redshift $z=20$, or even larger.  
\end{abstract}

\maketitle

\noindent{{\bf{\em Introduction.}}}
%
In the last decade a surprising connection between gravity in the strong field regime and particle physics 
has emerged in several contexts~\cite{Arvanitaki:2009fg,Barack:2018yly,Bertone:2019irm}. Probably the most spectacular one is the 
possibility to search for ultralight bosons with 
current~\cite{Arvanitaki:2009fg,Arvanitaki:2014wva,Arvanitaki:2016qwi,Brito:2017wnc,Brito:2017zvb,Palomba:2019vxe,Isi:2018pzk} and 
future~\cite{Audley:2017drz,Sathyaprakash:2019yqt,Baibhav:2019rsa,Sedda:2019uro} 
gravitational-wave~(GW) detectors.
Ultralight bosons (such as the QCD axion, axion-like particles, dark photons, etc) could be a significant component of 
the dark matter~\cite{Arvanitaki:2009fg,Essig:2013lka,Marsh:2015xka,Hui:2016ltb} and are predicted in a multitude of 
beyond Standard Model scenarios~\cite{Jaeckel:2010ni,Essig:2013lka,Hui:2016ltb,Irastorza:2018dyq}, including extra 
dimensions and string theories. They naturally interact very weekly and in a model-dependent fashion with baryonic 
matter, but their gravitational interaction is universal. 

A striking gravitational effect triggered by these fields near spinning black holes~(BHs) is the
superradiant instability~\cite{Press:1972zz,Detweiler:1980uk,Cardoso:2004nk,Brito:2015oca,Shlapentokh-Rothman:2013ysa}, 
which occurs whenever the 
boson frequency $\omega_R$ satisfies the superradiant condition $0<\omega_R<m \Omega_{\rm H}$,
where $\Omega_{\rm H}$ is the horizon angular velocity and $m$ is the
azimuthal quantum number of the unstable mode. 

Recent years have witnessed spectacular progress in understanding superradiant instabilities and their phenomenology, 
both for 
scalars~\cite{Damour:1976kh,Detweiler:1980uk,Zouros:1979iw,Dolan:2007mj,Arvanitaki:2014wva,Arvanitaki:2016qwi,
Brito:2017wnc,Brito:2017zvb} and for 
vectors~\cite{Pani:2012vp,Pani:2012bp,Witek:2012tr,Endlich:2016jgc,East:2017mrj,East:2017ovw,Baryakhtar:2017ngi, 
East:2018glu,Frolov:2018ezx,Dolan:2018dqv,Siemonsen:2019ebd}. 
In the superradiant regime 
the BH spins down, transferring energy and angular momentum to a mostly dipolar ($m=1$) boson
condensate until $\omega_R\sim \Omega_{\rm H}$. The condensate is then
dissipated through the emission of mostly quadrupolar GWs, with
frequency set by the boson mass $m_b\equiv \mu \hbar$ (we use $G=c=1$ units). On longer time scales this process 
continues for $m>1$ modes.
The mechanism is most effective when
the boson's Compton wavelength is comparable to the BH's gravitational radius, i.e. when the 
\emph{gravitational coupling} $\alpha\equiv M\mu={\cal O}(0.1)$, which requires $m_b\sim 
10^{-11}(M_\odot/M)\,{\rm eV}$~\cite{Brito:2015oca}.

Compared to the scalar and vector cases, very little is known about the much more involved problem of the superradiant 
instability triggered by massive tensor (i.e., spin-2) fields. To the best of our knowledge the only work on the 
subject performed a perturbative expansion to linear order in the spin~\cite{Brito:2013wya}, which is inaccurate in the 
most interesting regime of highly-spinning BHs. 
Furthermore, the coupling of a massive spin-2 field to gravity is highly 
nontrivial~\cite{Hinterbichler:2011tt,deRham:2010kj,Hassan:2011hr,Hassan:2011zd,deRham:2014zqa} and this increases the 
complexity of the problem.
In this work we fill a gap in the BH superradiance program by 
computing analytically for the first time the superradiant instability time scale and the GW emission from 
BH-condensates made of massive spin-2 fields. We work in the ``small-coupling'' limit, $\alpha\ll1$, but do not make any 
assumption on the BH spin.
As we shall argue, the phenomenology of the spin-2 superrandiant instability is similar to the spin-1 case, leading to 
exquisite constraints on beyond Standard Model tensor fields. Furthermore, novel effects occur in the spin-2 
case which are absent for scalars and vectors.

\noindent{{\bf{\em Massive spin-2 fields around spinning BHs.}}}
A massive tensor field cannot be trivially coupled to gravity~\cite{Hinterbichler:2011tt,deRham:2014zqa} and, at the 
nonlinear level, there is a unique way to couple two dynamical 
tensors~\cite{deRham:2010kj,Hassan:2011hr,Hassan:2011zd}.
On a curved, Ricci-flat\cite{deSitter}, 
spacetime $g_{ab}$, the unique action to cubic order in the spin-2 fields has been derived in 
Ref.~\cite{Babichev:2016bxi} and 
schematically reads
\begin{eqnarray}
 S^{(3)}=&&\int d^4x\sqrt{-g}\left.\Big[{\cal L}_{\rm GR}(G)+{\cal L}_{\rm GR}(H)\right.\nonumber\\
 &-&\left.\frac{\mu^2}{4}(H_{ab}H^{ab}-H^2)+{\cal L}_{\rm cubic}(G,H)\right]\,,  \label{action}
\end{eqnarray}
where $G_{ab}$ and $H_{ab}$ are the canonically normalized mass eigenstates describing a massless and a massive spin-2 
field, respectively, ${\cal L}_{\rm GR}$ is the 
Einstein-Hilbert Lagrangian truncated at quadratic order, ${\cal L}_{\rm cubic}$ (to be discussed in the Appendix) is a 
complicated interaction term that depends either linearly on $G_{ab}$ and quadratically on $H_{ab}$, or cubically on 
$G_{ab}$ and $H_{ab}$ independently.

To the zeroth order in the massive field $H_{ab}$, the field equations reduce to $R_{ab}(g)=0$ and we consistently assume a background Kerr metric, although our computation does not depend on the details of the 
background and should be valid also for different solutions that might exist in bimetric 
theories~\cite{Brito:2013xaa,Babichev:2015xha}. To first order, the linearized
field equations describing the five physical degrees of freedom of a massive spin-2 perturbation read~\cite{Brito:2013wya,Mazuet:2018ysa} 
\begin{eqnarray} 
	\Box H_{ab} + 2 R_{acbd} H^{cd} - \mu ^{2} H_{ab}&=&0	\label{eq:spin2_eom}\,, \\
	\nabla^{a} H_{ab}=0\,, \qquad {H^{a}}_{a}&=&0\,, \label{eq:spin2_constraints}
\end{eqnarray}
where the box operator, the Riemann tensor, and contractions are constructed with the background metric. 

\begin{figure}[t]
\includegraphics[width=0.51\textwidth]{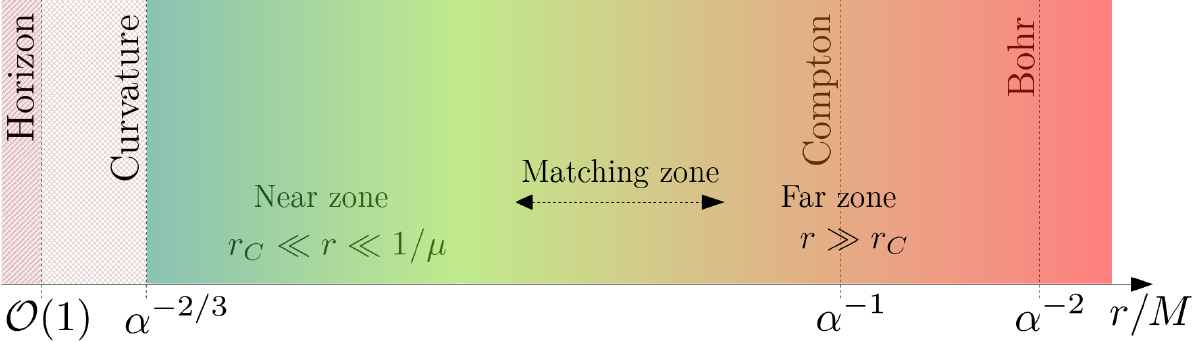}
\caption{Schematic representation of the length scales involved in the superradiant instability of massive 
spin-2 fields. In the small-coupling ($\alpha\equiv M\mu\ll1$) limit all scales are well separated from each other.
Our approximation fails to capture solutions with significant support within the curvature radius, $r_C= 
M/\alpha^{2/3}$. The standard superradiant modes instead have significant support around the Bohr radius, 
$r_{\rm Bohr}=M/\alpha^2$, the largest scale in the problem.
\label{fig:diagram}}
\end{figure}

Although Eqs.~\eqref{eq:spin2_eom} and \eqref{eq:spin2_constraints} are not separable on a Kerr background with 
standard methods, they can be solved in the $\alpha\ll1$ limit to compute the spectrum of unstable modes, as detailed 
in the Appendix.
Our method is based on matched asymptotics, namely the field equations are solved separately in a far zone ($r\gg 
M$) and in a near zone ($r\ll \mu^{-1}$). The two solutions can be matched in a common region when $\alpha\ll1$, see 
Fig.~\ref{fig:diagram}.

In addition, at variance with the scalar and vector cases, to be able to solve the field equations analytically we 
also need to consider the region where the Riemann tensor term in Eq.~\eqref{eq:spin2_eom} is much smaller than the 
mass term. This requires\footnote{Incidentally, $r_C$ coincides with the Vainshtein radius within which nonlinearities in massive gravity becomes important and allow recovering general relativity~\cite{deRham:2014zqa}. However, in our case this scale emerges already at the linear level.} 
\begin{equation}
 r\gg r_C\equiv M \alpha^{-2/3} \,. \label{Curvature}
\end{equation}
Since $M\ll r_C\ll 1/\mu$ in the small-coupling limit, the matching region satisfies the above condition.
Because of condition~\eqref{Curvature}, our method fails to capture eigenfunctions with significant support at 
$r\lesssim r_C$. This is the case for the unstable spherical mode that exists in the 
nonspinning case~\cite{Babichev:2013una,Brito:2013wya,Brito:2013xaa} and for the ``special'' dipole mode found numerically in 
Ref.~\cite{Brito:2013wya}\footnote{We explicitly checked 
that the spherical modes peak at $r\sim r_C$ whereas the special modes peak at $r<r_C$; both families of 
modes are absent if one neglects the Riemann tensor term in Eq.~\eqref{eq:spin2_eom}, i.e. they arise from the 
nontrivial coupling of the two tensor fields.}.  
On the other hand, as we will check a posteriori, the ordinary superradiant eigenfunctions have significant support only 
around the ``Bohr radius'', $r_{\rm Bohr}\sim M/\alpha^2$~\cite{Brito:2015oca}, and are therefore well reproduced by our 
analytical approximation.

In the far region the tensor $H_{ab}$ can be decomposed in a basis of ``pure-orbital'' tensor spherical 
harmonics~\cite{Mathews:1962,Thorne:1980ru,Maggiore:1900zz}. The radial dependence is entirely encoded in the hydrogen-like radial equation, whose 
eigenfunctions can be labeled by their orbital angular momentum $\ell\geq0$ and by 
the overtone number $n\geq0$ representing the number of nodes in the 
radial function. The energy levels are 
\begin{equation}
\omega_R\simeq\mu\left(1-\frac{\alpha^2}{2\left(\ell+n+1\right)^2}\right)\,, \label{wR}
\end{equation}
as in the scalar and vector cases. 

The instability time scale $\tau_{\rm inst}$ can be computed through the energy decay rate, $\Gamma=\dot 
E_H/M_c=1/\tau_{\rm 
inst}$, where $M_c$ is the energy of the condensate and $\dot E_H$ is the energy flux across the horizon, which can in 
turn be computed through the stress-energy tensor stemming from 
action~\eqref{action}~\cite{Babichev:2016bxi,Aoki:2017ixz}, and explicitly given in the Appendix.
%
%
By also making use of the BH absorption probability for long-wavelength massless spin-2 
waves~\cite{1973ZhETF..64...48S,1973ZhETF..65....3S,Brito:2015oca}, the computation detailed in the
Appendix yields 
\begin{equation}
 \Gamma = -C_{j\ell}\frac{{\cal P}_{jm}(\chi)}{{\cal P}_{jm}(0)}\alpha^{2(\ell+j)+5}(\omega_R-m\Omega_{\rm 
H})\,, \label{Gamma}
\end{equation}
where 
\begin{equation}
 {\cal 
P}_{jm}(\chi)=(1+\Delta)\Delta^{2j}\prod_{q=1}^j\left[1+4M^2\left(\frac{\omega_R-m\Omega_{\rm 
H}}{q\kappa}\right)^2\right]
\end{equation}
is proportional to the BH absorption probability, $\Delta=\sqrt{1-\chi^2}$, and $\kappa=\frac{\Delta}{1+\Delta}$.
The integer $j\in(|\ell-2|,\ell+2)\geq0$ is the total angular momentum, the integer $m\in(-j,j)$, and the constant 
$C_{j\ell}$ depends on the mode.
The superradiant instability requires a nonaxisymmetric mode ($m\neq0$) and therefore 
$j\geq1$. Hence, at variance with the scalar and vector cases, there exist \emph{two} dominant unstable modes with the same 
scaling $\tau_{\rm inst}\sim 
-\alpha^{-9}(\omega_R-m\Omega_{\rm H})^{-1}$: 
the \emph{dipole} $j=\ell=1$, and the \emph{quadrupole} $j=2$, $\ell=0$ (the latter being absent in the scalar and 
vector cases). We find $C_{20}=128/45$ and $C_{11}=10/9$, so that the quadrupole mode has always the 
shortest instability time scale.
When the BH spin is small the analytical results match very well the numerical ones obtained in Ref.~\cite{Brito:2013wya} to linear order in the spin.

\noindent{{\bf{\em GWs from spin-2 condensates around BHs.}}}
The GW dissipation time scale, $\tau_{\rm GW}$, can be computed from the stress-energy 
tensor of the condensate.
Crucially, $\tau_{\rm GW}\gg\tau_{\rm inst}\gg M$ 
in the small-coupling limit, so the process can be thought to 
occur in two stages~\cite{Brito:2014wla,East:2017ovw}.
In the first (linear) phase the condensate grows on a time scale given by $\tau_{\rm inst}=1/\Gamma$ for the most unstable modes [see 
Eq.~\eqref{Gamma}] until the superradiant condition $\omega_R\sim m\Omega_{\rm H}$ is
nearly saturated. In the second (nonlinear) phase GW emission governs the evolution of the condensate, which is
dissipated over the time scale $\tau_{\rm GW}\gg\tau_{\rm inst}$.
This separation of scales allows us to study the process in a quasi-adiabatic 
approximation~\cite{Brito:2014wla,East:2017ovw} using 
Teukolsky's formalism to compute the GW emission~\cite{Teukolsky:1973ha,Yoshino:2013ofa}. As detailed in the Appendix, in the small-frequency limit the GW flux $\dot{E}_{\rm GW}$ can be computed analytically~\cite{Poisson:1993vp,Brito:2014wla,Ficarra:2018rfu}.
For the most unstable modes we get 
\begin{equation}
 \tau_{\rm GW}\equiv \frac{ M_c}{\dot{E}_{\rm GW}} \sim D_{j\ell m} \frac{M^2}{M_c} \alpha^{-4\ell-10}
 \sim 290\left(\frac{0.2}{\alpha}\right)^{10}\,{\rm s}\,, \label{tauGW}
\end{equation}
where $M_c$ is the mass of the particular mode, $D_{202}\approx 2\times 10^{-2}$ and  $D_{111}\approx 2.6$ (their 
analytical form is given in the Appendix). As a useful estimate, in the 
last step we assumed $M_c\sim 0.1 M$~\cite{Brito:2014wla}, $M\sim 30\,{M_\odot}$, and $\ell=0$, showing that
$\tau_{\rm GW}\gg\tau_{\rm inst}$ can be relatively short.

The presence of two unstable modes with comparable time scales slightly complicates the above picture. The dipolar mode 
grows until $\Omega_{\rm H}=\omega_R$, reaching a mass $M_c^{m=1}$. At the same time, the BH spin keeps being extracted 
by the quadrupolar mode, which is still unstable in this regime and indeed saturates only at $\Omega_{\rm 
H}=\omega_R/2$, reaching a mass $M_c^{m=2}$. 
When the BH spins down such that $\Omega_{\rm H}<\omega_R$, the dipole mode leaves the superradiant regime and is 
quickly reabsorbed by the BH (since the absorption time scale is significantly shorter than that of GW emission), thus 
giving back almost all its mass and spin~\cite{Ficarra:2018rfu}. Therefore, at the time when $\Omega_{\rm 
H}=\omega_R/2$, the net mass and angular momentum loss are entirely due to the quadrupole mode, which is finally emitted 
in GWs over the time scale in Eq.~\eqref{tauGW}.

The emitted signal is nearly monochromatic, with frequency $f_s=\omega_R/\pi$, where $\omega_R$ is given in 
Eq.~\eqref{wR}. Thus, BH-boson condensates are continuous sources, like pulsars for LIGO/Virgo or verification binaries 
for LISA.
There are, however, two notable differences: (i) depending on the value of $\alpha$, the GW emission time scale 
$\tau_{\rm GW}$ can be significantly shorter than the observation time, resulting in an impulsive signal; (ii)~at 
variance with the case of massive scalar and vector fields, GW 
emission for the dominant spin-$2$ mode is mostly \emph{hexadecapolar}
and not quadrupolar, since it is produced by a spinning quadrupolar field and not by a spinning dipolar field. 
The hexadecapolar nature of the radiation implies that the signal vanishes along the BH spin axis, at 
variance with the quadrupolar case, for which it is maximum in that direction. 

To estimate the GW signal, we define the characteristic GW amplitude as
\begin{equation}
h_c=\sqrt{N_{\rm cycles}}\,h_{\rm rms}\,, 
\label{strain}
\end{equation}
where $N_{\rm cycles}\sim \min[\sqrt{f T_{\rm obs}},\sqrt{ f_s \tau_{\rm GW}}]$ is the approximate numbers of cycles in 
the detector,  $f=f_s/(1+z)$ is the detector frame frequency, $T_{\rm obs}$ is the observation time, and $h_{\rm rms} = \sqrt{\dot{E}_{\rm GW}/(5 f^2 r^2 
\pi^2)}$ is the root-mean-square amplitude obtained by averaging over source and detector orientations 
(see Refs.~\cite{Brito:2017wnc,Brito:2017zvb} for details).
\begin{figure*}[th]
\includegraphics[width=0.475\textwidth]{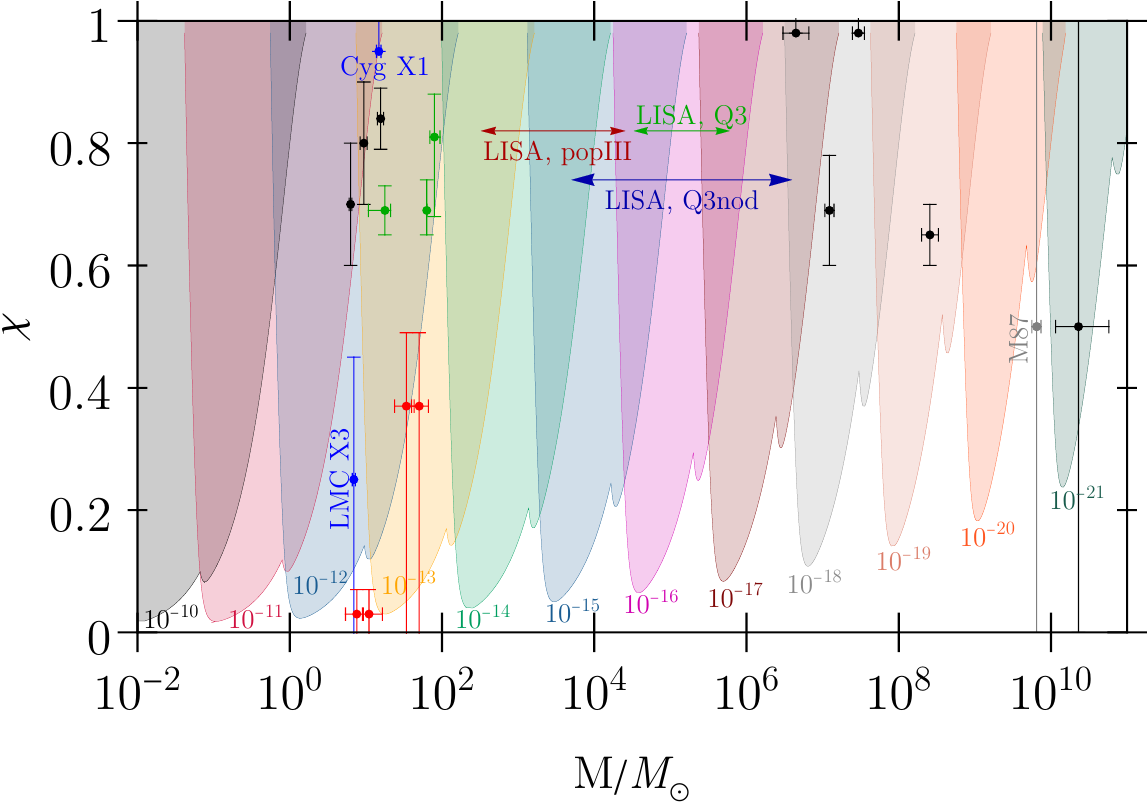}
\includegraphics[width=0.51\textwidth]{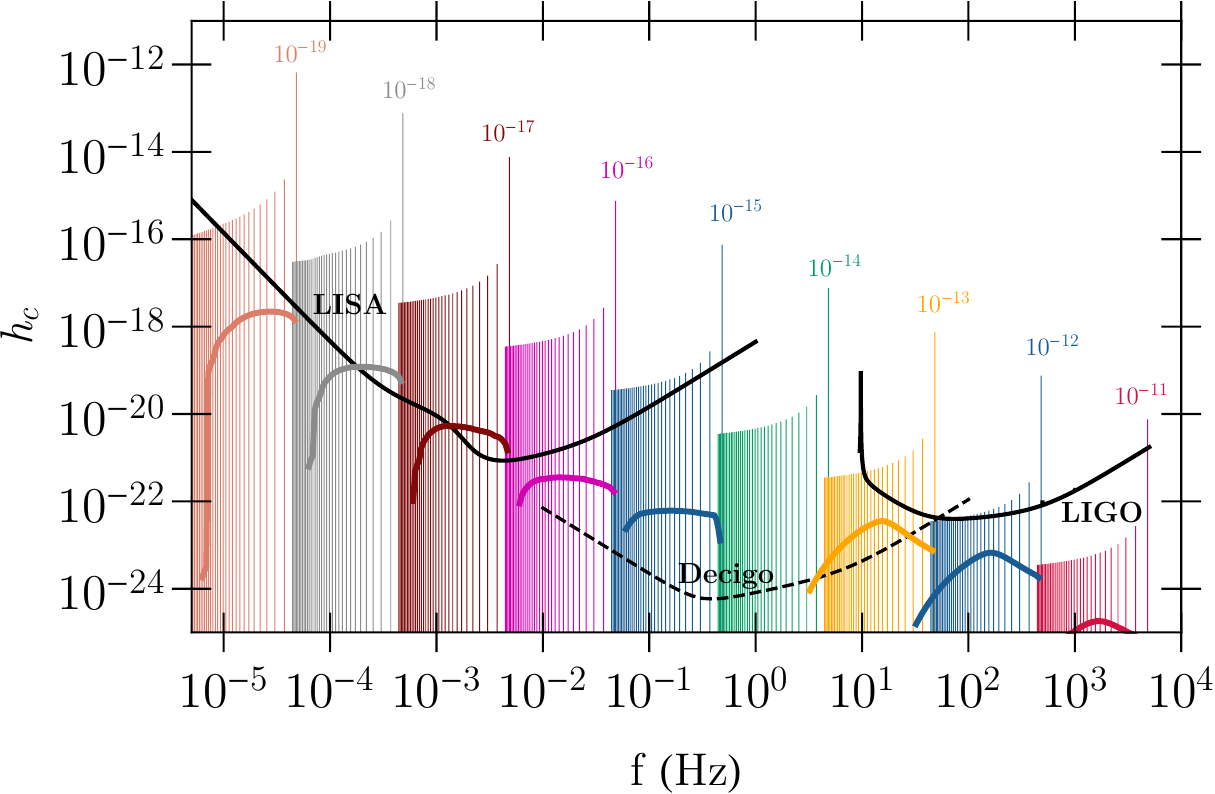}
\caption{\emph{Left:} Exclusion regions in the BH spin-mass diagram obtained from the superradiant 
instability of Kerr BHs against massive spin-2 fields for the most unstable quadrupolar ($j=2=m$, $\ell=0$) and 
octupolar ($j=3=m$, $\ell=1$) modes.
For each mass of the field (reported in units of eV), the separatrix corresponds to an instability time scale 
equal to the Salpeter time, $\tau_S=4.5\times 10^7 {\rm\, yr\,}$.
The meaning of the markers is explained in the main text.
\emph{Right:}
GW characteristic strain (thin lines) as defined by Eq.~\eqref{strain} for $T_{\rm obs}=4\,{\rm yr}$ produced by spin-2 
condensates compared to the characteristic noise strain of Advanced LIGO at design 
sensitivity~\cite{Aasi:2013wya} and to the sky-averaged characteristic noise strain of 
LISA~\cite{Audley:2017drz,Cornish:2018dyw} (black thick curves). 
The characteristic noise strain is defined as $\sqrt{f S_n(f)}$, with $S_n(f)$ being the noise power spectral density of 
the detector. Each (nearly vertical) line shows the strain for a 
given boson mass $m_b$, computed at redshift $z\in(0.001,10)$ (from right to left, in steps of $\delta z=0.3$), with 
$\alpha$ increasing in the superradiant range $(0,2M\Omega_{\rm H})$ along each line, and assuming initial BH spin 
$\chi_i=0.7$. Different colors correspond to different boson masses $m_b$. Thick colored lines show the stochastic
background produced by the whole population of astrophysical BHs under optimistic 
assumptions~\cite{Brito:2017wnc,Brito:2017zvb}, after subtracting the events that would be resolvable assuming $T_{\rm 
obs}=4\,{\rm yr}$ of coherent observation time.
The characteristic noise strain of DECIGO~\cite{Kawamura:2006up} (dashed line) is also shown for
reference.
\label{fig:Regge}}
\end{figure*}
%

\noindent{{\bf{\em Bounds from BH mass-spin distribution.}}}
We can now turn our attention to the phenomenology of the BH superradiant instability for massive spin-2 fields.
A generic prediction is that
highly spinning BHs would lose angular momentum over a time scale $\tau_{\rm inst}=1/\Gamma$ [see Eq.~\eqref{Gamma}] 
that might be much shorter than typical astrophysical time scales.
Thus, an indirect signature of ultralight bosons is statistical evidence for slowly rotating BHs in a part of the 
``Regge'' (mass versus angular momentum) plane of astrophysical 
BHs~\cite{Arvanitaki:2010sy,Brito:2015oca,Baryakhtar:2017ngi,Brito:2017wnc,Brito:2017zvb,Ng:2019jsx,Fernandez2019}. 

Our results are summarized in Fig.~\ref{fig:Regge}, whose left panel shows the ``forbidden'' regions in the 
Regge plane for selected values of $m_b$, obtained by requiring that the 
instability acts on time scales shorter than known ``spin-up'' astrophysical processes such as accretion. Here we 
conservatively require that $\tau_{\rm inst}$ be shorter than the Salpeter time scale for accretion, $\tau_S=4.5\times 
10^7 {\rm\, yr\,}$.
Data points (with error bars) in the left panel of Fig.~\ref{fig:Regge} refer to different observations:
(i)~Black points denote electromagnetic estimates of 
stellar and supermassive BH spins obtained using either the K$\alpha$ iron line or the continuum fitting
method~\cite{Brenneman:2011wz,Middleton:2015osa}.
(ii)~Red points are the $90\%$ confidence levels for the spins of the
primary and secondary BHs in (a selection of) the merger events detected in LIGO-Virgo first two 
runs~\cite{LIGOScientific:2018mvr,Venumadhav:2019lyq}. 
Here we use the errors on $\chi_{\rm eff}\equiv \frac{m_1\chi_1+m_2\chi_2}{m_1+m_2}$ as a proxy for the errors on the 
individual spins, $\chi_1$ and $\chi_2$. Whilst the binary spins measured so far with GWs are 
affected by large uncertainties and are anyway compatible to zero for almost all sources (but 
see~\cite{LIGOScientific:2018mvr,Venumadhav:2019lyq} for a few events in which $\chi_{\rm eff}\neq0$), future 
detections will provide measurements of the individual spins with $30\%$ accuracy~\cite{TheLIGOScientific:2016pea}. 
(iii)~Green points are the $90\%$ confidence levels for the mass-spin of a selection of the GW coalescence 
remnants~\cite{LIGOScientific:2018mvr}. While those events cannot be used to constrain the Regge plane (because the 
observation time scale is much shorter than $\tau_{\rm inst}$), they identify targets of merger follow-up 
searches~\cite{Arvanitaki:2014wva,Arvanitaki:2016qwi,Baryakhtar:2017ngi,Isi:2018pzk,Ghosh:2018gaw}. This is particularly 
important in the spin-$2$ case, where $\tau_{\rm inst}$ can be as small as a fraction of seconds for typical remnants in 
the LIGO/Virgo band.
(iv) 
Instead of using $\tau_S$ as a reference time scale, more direct constraints would come from comparing $\tau_{\rm 
inst}$ 
against the baseline (typically ${\cal O}(10\,{\rm yr})$) during which the spin of certain BH candidates is 
measured to be constant~\cite{Cardoso:2018tly}, as it is the case for LMC X-3~\cite{Steiner:2010kd} and 
Cyg~X-1~\cite{Gou:2009ks}, shown in the left panel of Fig.~\ref{fig:Regge} by blue points.
In particular, Cyg~X-1 can confidently exclude the range $2.9\times10^{-13}<m_b/{\rm eV}<9.8\times10^{-12}$.
(v)~Finally, the single gray point is the mass of M87 measured by the Event Horizon 
Telescope~\cite{Akiyama:2019cqa,Akiyama:2019eap}. While a 
direct spin measurement is still not available, M87 has been suggested to have a large 
spin~\cite{Akiyama:2019fyp,Tamburini:2019vrf}. A putative measurement $\chi_{\rm M87}\gtrsim0.2$ would constrain the 
mass range $m_b\sim 10^{-20}$--$10^{-21}$~eV~\cite{Davoudiasl:2019nlo,Chen:2019fsq,Cunha:2019ikd}.
If the largest known supermassive BHs with $M\simeq 2\times 10^{10} 
M_\odot$ \cite{McConnell:2011mu,2012arXiv1203.1620M} were confirmed to have nonzero 
spin~\cite{Riechers:2008xt}, we could get even more stringent bounds.

Very precise spin measurements of binary BH components out to cosmological distances will come from the future LISA 
mission~\cite{Audley:2017drz}. 
Depending on the mass of BH seeds in the early Universe, LISA will also detect intermediate mass BHs, thus probing the 
existence of ultralight bosons in a large mass range (roughly $m_b\sim 10^{-14}$--$10^{-17}$~eV)
that is inaccessible to electromagnetic observations of stellar and
supermassive BHs and to ground-based GW detectors~\cite{Brito:2017wnc,Brito:2017zvb,Cardoso:2018tly}.
This is shown in the left panel of Fig.~\ref{fig:Regge} by the horizonal arrows, which denote the range of projected 
LISA measurements using three different population models for supermassive BH growth~\cite{Klein:2015hvg,Brito:2017zvb}.

Owing to the wideness of the Regge gaps, the range of detectable spin-$2$ masses is larger than in 
the scalar case and similar to the vector case. If (spinning) BHs of a few solar masses are 
detected~\cite{Kyutoku:2020xka}, they can probe $m_b\sim 10^{-10}\,{\rm eV}$, whereas BHs as massive as M87 can reach 
the other hand of the spectrum, $m_b\sim 10^{-21}\,{\rm eV}$, where ultralight bosons are also compelling dark-matter 
candidates~\cite{Hui:2016ltb}.

\begin{figure}[th]
\includegraphics[width=0.49\textwidth]{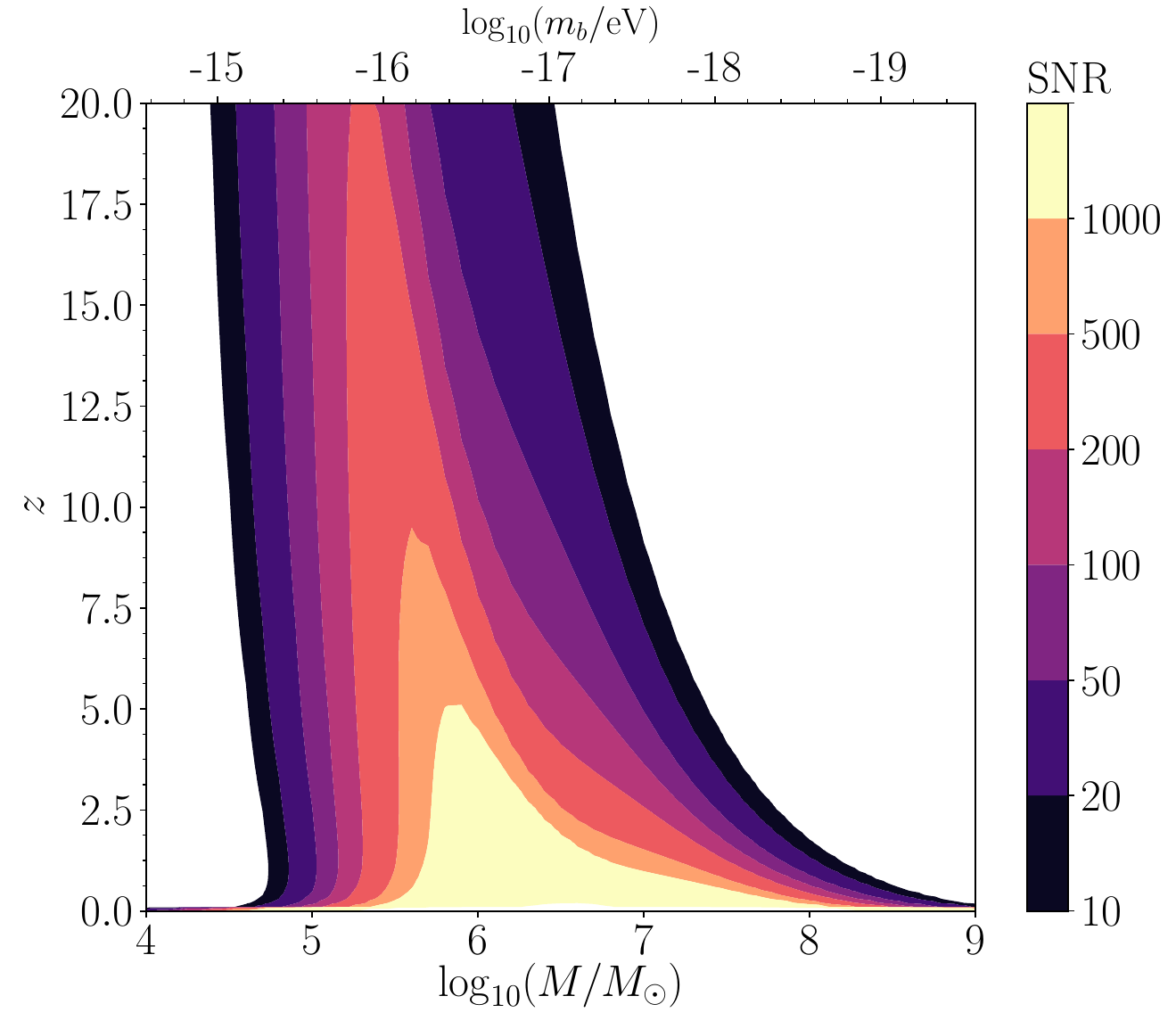}
\caption{Waterfall plot~\cite{Audley:2017drz} showing the angle-averaged LISA SNR of continuous GWs from massive 
spin-2 condensates for different source-frame BH masses and at different redshift. For simplicity we 
assumed $\alpha=0.2$, $\chi=0.7$, and neglected the 
confusion noise from the stochastic background (see Fig.~\ref{fig:Regge}). The SNR is approximately given by ${\rm 
SNR}\sim 
\sqrt{5} h_c/\sqrt{f S_n(f)}$ (where the factor $\sqrt{5}$ comes from the fact that we are using a sky-averaged LISA 
power spectral density, see e.g. ~\cite{Berti:2005ys}) and we assumed $T_{\rm obs}=4 {\rm yr}$.
\label{fig:waterfall}}
\end{figure}
%

\noindent{{\bf{\em Direct GW signatures.}}}
In the right panel of Fig.~\ref{fig:Regge} we compare the GW characteristic strain of Eq.~\eqref{strain} with the 
characteristic noise strain of current and future GW detectors. For a given boson mass $m_b$ and redshift $z$, the GW 
frequency depends very weakly on $\alpha$, whereas the GW strain is maximum for couplings near the superradiant 
threshold, $\alpha\lesssim 2M\Omega_{\rm H}$.
Each point in the (nearly vertical) lines corresponds to a single source with a given $\alpha$ at different redshift 
$z\in(0.001,10)$. 
Interestingly --~owing to the redshift of the frequency~-- sources at high redshift can emit in the optimal frequency 
bucket even when their frequency at $z\sim0$ is marginally detectable.
Furthermore, in the LISA band the signal at 
high redshift decreases more slowly than the slope of the noise, allowing to potentially detect sources at cosmological 
distances. This is better shown in the ``waterfall''~\cite{Audley:2017drz} plot in Fig.~\ref{fig:waterfall} where we 
show a typical angle-averaged redshift horizon for LISA.
Remarkably, the continuous GW signal could be detected even when $z\approx20$ or higher: every 
supermassive BH in the universe with masses $10^{4.5}\lesssim M/M_\odot\lesssim 10^{6.5}$ 
can potentially be a detectable source if the boson mass is in the optimal range.

Note that LIGO can potentially probe a larger range of spin-$2$ masses than current bounds from the mass-spin 
measurements of stellar mass BHs, although the rates for direct GW detections~\cite{Brito:2017zvb} which can provide 
more stringent constraints depend on the formation rate of spinning BHs with masses $M\gtrsim 20M_\odot$.

Finally, thick solid curves in the right panel of Fig.~\ref{fig:Regge} correspond to the 
stochastic background from the whole BH population, for a boson mass $m_b$, computed with the same technique as in 
Refs.~\cite{Brito:2017wnc,Brito:2017zvb} and assuming the optimistic BH mass and spin distributions of 
Refs.~\cite{Brito:2017wnc,Brito:2017zvb}. Roughly speaking, when the stochastic signal is higher than the detector's 
noise curve, it produces a ``confusion noise'' which can complicate the detection of individual 
sources~\cite{Brito:2017wnc,Brito:2017zvb}. 
In the most optimistic scenario the background would be observable by Advanced LIGO and LISA in the ranges $\sim 
[5\times 10^{-14},10^{-12}]$ eV and $\sim 5\times [10^{-19},10^{-16}]$ eV, respectively.  

To summarize, spin-2 fields with masses
$10^{-19}\,{\rm eV} \lesssim m_b \lesssim 10^{-11}\,{\rm eV}$ (with a
small gap around $m_b\sim 10^{-14}$~eV, which might be filled by
DECIGO~\cite{Kawamura:2006up}) would turn BHs into exotic sources of continuous GWs and of a stochastic background 
detectable by GW detectors up to cosmological distances.

Our results can be implemented in direct search pipelines for continuous sources, along the lines of axion 
searches~\cite{Tsukada:2018mbp,Palomba:2019vxe,Sun:2019mqb,Zhu:2020tht}.
Compared to the scalar case, the frequency drift for spin-$2$ clouds is much faster since $\dot f_s \propto \dot E_{\rm 
GW}\sim \alpha^{4\ell+10}$ is a factor ${\cal O}(\alpha^{-4})$ larger than in the scalar case. 
Thus, spin-$2$ direct searches can be implemented with the same techniques as in the spin-$1$ 
case~\cite{Baryakhtar:2017ngi}.
In addition, other detection strategies will include follow-up searches of post-merger remnants~\cite{Arvanitaki:2016qwi,Baryakhtar:2017ngi,Isi:2018pzk,Ghosh:2018gaw}, and self-gravity~\cite{Hannuksela:2018izj} and tidal effects~\cite{Baumann:2018vus,Zhang:2018kib,Berti:2019wnn,Baumann:2019eav,Baumann:2019ztm,Zhang:2019eid,Cardoso:2020hca}
of the condensate in BH binary inspirals. These will require an independent study of the full spectrum of the condensate 
that is left for the future.

\noindent{{\bf{\em Acknowledgments.}}}
We thank Eugeny Babichev for interesting discussion.
R.B. acknowledges financial support from the European Union's Horizon 2020 research and innovation programme under the 
Marie Sk\l odowska-Curie grant agreement No. 792862.
P.P. acknowledges financial support provided under the European Union's H2020 ERC, Starting 
Grant agreement no.~DarkGRA--757480, and , and under the MIUR PRIN and FARE programmes (GW-NEXT, CUP:~B84I20000100001).
The authors would like to acknowledge networking support by the COST Action CA16104 and 
support from the Amaldi Research Center funded by the MIUR program "Dipartimento di 
Eccellenza" (CUP:~B81I18001170001). 
This research was supported by the Munich Institute for Astro- and Particle Physics~(MIAPP) which is funded by the 
Deutsche Forschungsgemeinschaft (DFG, German Research Foundation) under Germany's Excellence Strategy - EXC-2094 - 
390783311.

\bibliography{notes}

\begin{thebibliography}{102}%
\makeatletter
\providecommand \@ifxundefined [1]{%
 \@ifx{#1\undefined}
}%
\providecommand \@ifnum [1]{%
 \ifnum #1\expandafter \@firstoftwo
 \else \expandafter \@secondoftwo
 \fi
}%
\providecommand \@ifx [1]{%
 \ifx #1\expandafter \@firstoftwo
 \else \expandafter \@secondoftwo
 \fi
}%
\providecommand \natexlab [1]{#1}%
\providecommand \enquote  [1]{``#1''}%
\providecommand \bibnamefont  [1]{#1}%
\providecommand \bibfnamefont [1]{#1}%
\providecommand \citenamefont [1]{#1}%
\providecommand \href@noop [0]{\@secondoftwo}%
\providecommand \href [0]{\begingroup \@sanitize@url \@href}%
\providecommand \@href[1]{\@@startlink{#1}\@@href}%
\providecommand \@@href[1]{\endgroup#1\@@endlink}%
\providecommand \@sanitize@url [0]{\catcode `\\12\catcode `\$12\catcode
  `\&12\catcode `\#12\catcode `\^12\catcode `\_12\catcode `\%12\relax}%
\providecommand \@@startlink[1]{}%
\providecommand \@@endlink[0]{}%
\providecommand \url  [0]{\begingroup\@sanitize@url \@url }%
\providecommand \@url [1]{\endgroup\@href {#1}{\urlprefix }}%
\providecommand \urlprefix  [0]{URL }%
\providecommand \Eprint [0]{\href }%
\providecommand \doibase [0]{http://dx.doi.org/}%
\providecommand \selectlanguage [0]{\@gobble}%
\providecommand \bibinfo  [0]{\@secondoftwo}%
\providecommand \bibfield  [0]{\@secondoftwo}%
\providecommand \translation [1]{[#1]}%
\providecommand \BibitemOpen [0]{}%
\providecommand \bibitemStop [0]{}%
\providecommand \bibitemNoStop [0]{.\EOS\space}%
\providecommand \EOS [0]{\spacefactor3000\relax}%
\providecommand \BibitemShut  [1]{\csname bibitem#1\endcsname}%
\let\auto@bib@innerbib\@empty
\bibitem [{\citenamefont {Arvanitaki}\ \emph {et~al.}(2010)\citenamefont
  {Arvanitaki}, \citenamefont {Dimopoulos}, \citenamefont {Dubovsky},
  \citenamefont {Kaloper},\ and\ \citenamefont
  {March-Russell}}]{Arvanitaki:2009fg}%
  \BibitemOpen
  \bibfield  {author} {\bibinfo {author} {\bibfnamefont {A.}~\bibnamefont
  {Arvanitaki}}, \bibinfo {author} {\bibfnamefont {S.}~\bibnamefont
  {Dimopoulos}}, \bibinfo {author} {\bibfnamefont {S.}~\bibnamefont
  {Dubovsky}}, \bibinfo {author} {\bibfnamefont {N.}~\bibnamefont {Kaloper}}, \
  and\ \bibinfo {author} {\bibfnamefont {J.}~\bibnamefont {March-Russell}},\
  }\href {\doibase 10.1103/PhysRevD.81.123530} {\bibfield  {journal} {\bibinfo
  {journal} {Phys.Rev.}\ }\textbf {\bibinfo {volume} {D81}},\ \bibinfo {pages}
  {123530} (\bibinfo {year} {2010})},\ \Eprint {http://arxiv.org/abs/0905.4720}
  {arXiv:0905.4720 [hep-th]} \BibitemShut {NoStop}%
\bibitem [{\citenamefont {Barack}\ \emph {et~al.}(2019)\citenamefont {Barack}
  \emph {et~al.}}]{Barack:2018yly}%
  \BibitemOpen
  \bibfield  {author} {\bibinfo {author} {\bibfnamefont {L.}~\bibnamefont
  {Barack}} \emph {et~al.},\ }\href {\doibase 10.1088/1361-6382/ab0587}
  {\bibfield  {journal} {\bibinfo  {journal} {Class. Quant. Grav.}\ }\textbf
  {\bibinfo {volume} {36}},\ \bibinfo {pages} {143001} (\bibinfo {year}
  {2019})},\ \Eprint {http://arxiv.org/abs/1806.05195} {arXiv:1806.05195
  [gr-qc]} \BibitemShut {NoStop}%
\bibitem [{\citenamefont {Bertone}\ \emph {et~al.}(2019)\citenamefont {Bertone}
  \emph {et~al.}}]{Bertone:2019irm}%
  \BibitemOpen
  \bibfield  {author} {\bibinfo {author} {\bibfnamefont {G.}~\bibnamefont
  {Bertone}} \emph {et~al.},\ }\href@noop {} {\  (\bibinfo {year} {2019})},\
  \Eprint {http://arxiv.org/abs/1907.10610} {arXiv:1907.10610 [astro-ph.CO]}
  \BibitemShut {NoStop}%
\bibitem [{\citenamefont {Arvanitaki}\ \emph {et~al.}(2015)\citenamefont
  {Arvanitaki}, \citenamefont {Baryakhtar},\ and\ \citenamefont
  {Huang}}]{Arvanitaki:2014wva}%
  \BibitemOpen
  \bibfield  {author} {\bibinfo {author} {\bibfnamefont {A.}~\bibnamefont
  {Arvanitaki}}, \bibinfo {author} {\bibfnamefont {M.}~\bibnamefont
  {Baryakhtar}}, \ and\ \bibinfo {author} {\bibfnamefont {X.}~\bibnamefont
  {Huang}},\ }\href {\doibase 10.1103/PhysRevD.91.084011} {\bibfield  {journal}
  {\bibinfo  {journal} {Phys. Rev.}\ }\textbf {\bibinfo {volume} {D91}},\
  \bibinfo {pages} {084011} (\bibinfo {year} {2015})},\ \Eprint
  {http://arxiv.org/abs/1411.2263} {arXiv:1411.2263 [hep-ph]} \BibitemShut
  {NoStop}%
\bibitem [{\citenamefont {Arvanitaki}\ \emph {et~al.}(2017)\citenamefont
  {Arvanitaki}, \citenamefont {Baryakhtar}, \citenamefont {Dimopoulos},
  \citenamefont {Dubovsky},\ and\ \citenamefont
  {Lasenby}}]{Arvanitaki:2016qwi}%
  \BibitemOpen
  \bibfield  {author} {\bibinfo {author} {\bibfnamefont {A.}~\bibnamefont
  {Arvanitaki}}, \bibinfo {author} {\bibfnamefont {M.}~\bibnamefont
  {Baryakhtar}}, \bibinfo {author} {\bibfnamefont {S.}~\bibnamefont
  {Dimopoulos}}, \bibinfo {author} {\bibfnamefont {S.}~\bibnamefont
  {Dubovsky}}, \ and\ \bibinfo {author} {\bibfnamefont {R.}~\bibnamefont
  {Lasenby}},\ }\href {\doibase 10.1103/PhysRevD.95.043001} {\bibfield
  {journal} {\bibinfo  {journal} {Phys. Rev.}\ }\textbf {\bibinfo {volume}
  {D95}},\ \bibinfo {pages} {043001} (\bibinfo {year} {2017})},\ \Eprint
  {http://arxiv.org/abs/1604.03958} {arXiv:1604.03958 [hep-ph]} \BibitemShut
  {NoStop}%
\bibitem [{\citenamefont {Brito}\ \emph
  {et~al.}(2017{\natexlab{a}})\citenamefont {Brito}, \citenamefont {Ghosh},
  \citenamefont {Barausse}, \citenamefont {Berti}, \citenamefont {Cardoso},
  \citenamefont {Dvorkin}, \citenamefont {Klein},\ and\ \citenamefont
  {Pani}}]{Brito:2017wnc}%
  \BibitemOpen
  \bibfield  {author} {\bibinfo {author} {\bibfnamefont {R.}~\bibnamefont
  {Brito}}, \bibinfo {author} {\bibfnamefont {S.}~\bibnamefont {Ghosh}},
  \bibinfo {author} {\bibfnamefont {E.}~\bibnamefont {Barausse}}, \bibinfo
  {author} {\bibfnamefont {E.}~\bibnamefont {Berti}}, \bibinfo {author}
  {\bibfnamefont {V.}~\bibnamefont {Cardoso}}, \bibinfo {author} {\bibfnamefont
  {I.}~\bibnamefont {Dvorkin}}, \bibinfo {author} {\bibfnamefont
  {A.}~\bibnamefont {Klein}}, \ and\ \bibinfo {author} {\bibfnamefont
  {P.}~\bibnamefont {Pani}},\ }\href {\doibase 10.1103/PhysRevLett.119.131101}
  {\bibfield  {journal} {\bibinfo  {journal} {Phys. Rev. Lett.}\ }\textbf
  {\bibinfo {volume} {119}},\ \bibinfo {pages} {131101} (\bibinfo {year}
  {2017}{\natexlab{a}})},\ \Eprint {http://arxiv.org/abs/1706.05097}
  {arXiv:1706.05097 [gr-qc]} \BibitemShut {NoStop}%
\bibitem [{\citenamefont {Brito}\ \emph
  {et~al.}(2017{\natexlab{b}})\citenamefont {Brito}, \citenamefont {Ghosh},
  \citenamefont {Barausse}, \citenamefont {Berti}, \citenamefont {Cardoso},
  \citenamefont {Dvorkin}, \citenamefont {Klein},\ and\ \citenamefont
  {Pani}}]{Brito:2017zvb}%
  \BibitemOpen
  \bibfield  {author} {\bibinfo {author} {\bibfnamefont {R.}~\bibnamefont
  {Brito}}, \bibinfo {author} {\bibfnamefont {S.}~\bibnamefont {Ghosh}},
  \bibinfo {author} {\bibfnamefont {E.}~\bibnamefont {Barausse}}, \bibinfo
  {author} {\bibfnamefont {E.}~\bibnamefont {Berti}}, \bibinfo {author}
  {\bibfnamefont {V.}~\bibnamefont {Cardoso}}, \bibinfo {author} {\bibfnamefont
  {I.}~\bibnamefont {Dvorkin}}, \bibinfo {author} {\bibfnamefont
  {A.}~\bibnamefont {Klein}}, \ and\ \bibinfo {author} {\bibfnamefont
  {P.}~\bibnamefont {Pani}},\ }\href {\doibase 10.1103/PhysRevD.96.064050}
  {\bibfield  {journal} {\bibinfo  {journal} {Phys. Rev.}\ }\textbf {\bibinfo
  {volume} {D96}},\ \bibinfo {pages} {064050} (\bibinfo {year}
  {2017}{\natexlab{b}})},\ \Eprint {http://arxiv.org/abs/1706.06311}
  {arXiv:1706.06311 [gr-qc]} \BibitemShut {NoStop}%
\bibitem [{\citenamefont {Palomba}\ \emph {et~al.}(2019)\citenamefont {Palomba}
  \emph {et~al.}}]{Palomba:2019vxe}%
  \BibitemOpen
  \bibfield  {author} {\bibinfo {author} {\bibfnamefont {C.}~\bibnamefont
  {Palomba}} \emph {et~al.},\ }\href {\doibase 10.1103/PhysRevLett.123.171101}
  {\bibfield  {journal} {\bibinfo  {journal} {Phys. Rev. Lett.}\ }\textbf
  {\bibinfo {volume} {123}},\ \bibinfo {pages} {171101} (\bibinfo {year}
  {2019})},\ \Eprint {http://arxiv.org/abs/1909.08854} {arXiv:1909.08854
  [astro-ph.HE]} \BibitemShut {NoStop}%
\bibitem [{\citenamefont {Isi}\ \emph {et~al.}(2019)\citenamefont {Isi},
  \citenamefont {Sun}, \citenamefont {Brito},\ and\ \citenamefont
  {Melatos}}]{Isi:2018pzk}%
  \BibitemOpen
  \bibfield  {author} {\bibinfo {author} {\bibfnamefont {M.}~\bibnamefont
  {Isi}}, \bibinfo {author} {\bibfnamefont {L.}~\bibnamefont {Sun}}, \bibinfo
  {author} {\bibfnamefont {R.}~\bibnamefont {Brito}}, \ and\ \bibinfo {author}
  {\bibfnamefont {A.}~\bibnamefont {Melatos}},\ }\href {\doibase
  10.1103/PhysRevD.99.084042} {\bibfield  {journal} {\bibinfo  {journal} {Phys.
  Rev.}\ }\textbf {\bibinfo {volume} {D99}},\ \bibinfo {pages} {084042}
  (\bibinfo {year} {2019})},\ \Eprint {http://arxiv.org/abs/1810.03812}
  {arXiv:1810.03812 [gr-qc]} \BibitemShut {NoStop}%
\bibitem [{\citenamefont {Audley}\ \emph {et~al.}(2017)\citenamefont {Audley}
  \emph {et~al.}}]{Audley:2017drz}%
  \BibitemOpen
  \bibfield  {author} {\bibinfo {author} {\bibfnamefont {H.}~\bibnamefont
  {Audley}} \emph {et~al.} (\bibinfo {collaboration} {LISA}),\ }\href@noop {}
  {\  (\bibinfo {year} {2017})},\ \Eprint {http://arxiv.org/abs/1702.00786}
  {arXiv:1702.00786 [astro-ph.IM]} \BibitemShut {NoStop}%
\bibitem [{\citenamefont {Sathyaprakash}\ \emph {et~al.}(2019)\citenamefont
  {Sathyaprakash} \emph {et~al.}}]{Sathyaprakash:2019yqt}%
  \BibitemOpen
  \bibfield  {author} {\bibinfo {author} {\bibfnamefont {B.~S.}\ \bibnamefont
  {Sathyaprakash}} \emph {et~al.},\ }\href@noop {} {\  (\bibinfo {year}
  {2019})},\ \Eprint {http://arxiv.org/abs/1903.09221} {arXiv:1903.09221
  [astro-ph.HE]} \BibitemShut {NoStop}%
\bibitem [{\citenamefont {Baibhav}\ \emph {et~al.}(2019)\citenamefont {Baibhav}
  \emph {et~al.}}]{Baibhav:2019rsa}%
  \BibitemOpen
  \bibfield  {author} {\bibinfo {author} {\bibfnamefont {V.}~\bibnamefont
  {Baibhav}} \emph {et~al.},\ }\href@noop {} {\  (\bibinfo {year} {2019})},\
  \Eprint {http://arxiv.org/abs/1908.11390} {arXiv:1908.11390 [astro-ph.HE]}
  \BibitemShut {NoStop}%
\bibitem [{\citenamefont {Sedda}\ \emph {et~al.}(2019)\citenamefont {Sedda}
  \emph {et~al.}}]{Sedda:2019uro}%
  \BibitemOpen
  \bibfield  {author} {\bibinfo {author} {\bibfnamefont {M.~A.}\ \bibnamefont
  {Sedda}} \emph {et~al.},\ }\href@noop {} {\  (\bibinfo {year} {2019})},\
  \Eprint {http://arxiv.org/abs/1908.11375} {arXiv:1908.11375 [gr-qc]}
  \BibitemShut {NoStop}%
\bibitem [{\citenamefont {Essig}\ \emph {et~al.}(2013)\citenamefont {Essig}
  \emph {et~al.}}]{Essig:2013lka}%
  \BibitemOpen
  \bibfield  {author} {\bibinfo {author} {\bibfnamefont {R.}~\bibnamefont
  {Essig}} \emph {et~al.},\ }in\ \href
  {http://inspirehep.net/record/1263039/files/arXiv:1311.0029.pdf} {\emph
  {\bibinfo {booktitle} {{Community Summer Study 2013: Snowmass on the
  Mississippi (CSS2013) Minneapolis, MN, USA, July 29-August 6, 2013}}}}\
  (\bibinfo {year} {2013})\ \Eprint {http://arxiv.org/abs/1311.0029}
  {arXiv:1311.0029 [hep-ph]} \BibitemShut {NoStop}%
\bibitem [{\citenamefont {Marsh}(2016)}]{Marsh:2015xka}%
  \BibitemOpen
  \bibfield  {author} {\bibinfo {author} {\bibfnamefont {D.~J.~E.}\
  \bibnamefont {Marsh}},\ }\href {\doibase 10.1016/j.physrep.2016.06.005}
  {\bibfield  {journal} {\bibinfo  {journal} {Phys. Rept.}\ }\textbf {\bibinfo
  {volume} {643}},\ \bibinfo {pages} {1} (\bibinfo {year} {2016})},\ \Eprint
  {http://arxiv.org/abs/1510.07633} {arXiv:1510.07633 [astro-ph.CO]}
  \BibitemShut {NoStop}%
\bibitem [{\citenamefont {Hui}\ \emph {et~al.}(2017)\citenamefont {Hui},
  \citenamefont {Ostriker}, \citenamefont {Tremaine},\ and\ \citenamefont
  {Witten}}]{Hui:2016ltb}%
  \BibitemOpen
  \bibfield  {author} {\bibinfo {author} {\bibfnamefont {L.}~\bibnamefont
  {Hui}}, \bibinfo {author} {\bibfnamefont {J.~P.}\ \bibnamefont {Ostriker}},
  \bibinfo {author} {\bibfnamefont {S.}~\bibnamefont {Tremaine}}, \ and\
  \bibinfo {author} {\bibfnamefont {E.}~\bibnamefont {Witten}},\ }\href
  {\doibase 10.1103/PhysRevD.95.043541} {\bibfield  {journal} {\bibinfo
  {journal} {Phys. Rev.}\ }\textbf {\bibinfo {volume} {D95}},\ \bibinfo {pages}
  {043541} (\bibinfo {year} {2017})},\ \Eprint
  {http://arxiv.org/abs/1610.08297} {arXiv:1610.08297 [astro-ph.CO]}
  \BibitemShut {NoStop}%
\bibitem [{\citenamefont {Jaeckel}\ and\ \citenamefont
  {Ringwald}(2010)}]{Jaeckel:2010ni}%
  \BibitemOpen
  \bibfield  {author} {\bibinfo {author} {\bibfnamefont {J.}~\bibnamefont
  {Jaeckel}}\ and\ \bibinfo {author} {\bibfnamefont {A.}~\bibnamefont
  {Ringwald}},\ }\href {\doibase 10.1146/annurev.nucl.012809.104433} {\bibfield
   {journal} {\bibinfo  {journal} {Ann. Rev. Nucl. Part. Sci.}\ }\textbf
  {\bibinfo {volume} {60}},\ \bibinfo {pages} {405} (\bibinfo {year} {2010})},\
  \Eprint {http://arxiv.org/abs/1002.0329} {arXiv:1002.0329 [hep-ph]}
  \BibitemShut {NoStop}%
\bibitem [{\citenamefont {Irastorza}\ and\ \citenamefont
  {Redondo}(2018)}]{Irastorza:2018dyq}%
  \BibitemOpen
  \bibfield  {author} {\bibinfo {author} {\bibfnamefont {I.~G.}\ \bibnamefont
  {Irastorza}}\ and\ \bibinfo {author} {\bibfnamefont {J.}~\bibnamefont
  {Redondo}},\ }\href {\doibase 10.1016/j.ppnp.2018.05.003} {\bibfield
  {journal} {\bibinfo  {journal} {Prog.Part.Nucl.Phys.}\ }\textbf {\bibinfo
  {volume} {102}},\ \bibinfo {pages} {89} (\bibinfo {year} {2018})},\ \Eprint
  {http://arxiv.org/abs/1801.08127} {arXiv:1801.08127 [hep-ph]} \BibitemShut
  {NoStop}%
\bibitem [{\citenamefont {Press}\ and\ \citenamefont
  {Teukolsky}(1972)}]{Press:1972zz}%
  \BibitemOpen
  \bibfield  {author} {\bibinfo {author} {\bibfnamefont {W.~H.}\ \bibnamefont
  {Press}}\ and\ \bibinfo {author} {\bibfnamefont {S.~A.}\ \bibnamefont
  {Teukolsky}},\ }\href {\doibase 10.1038/238211a0} {\bibfield  {journal}
  {\bibinfo  {journal} {Nature}\ }\textbf {\bibinfo {volume} {238}},\ \bibinfo
  {pages} {211} (\bibinfo {year} {1972})}\BibitemShut {NoStop}%
\bibitem [{\citenamefont {Detweiler}(1980)}]{Detweiler:1980uk}%
  \BibitemOpen
  \bibfield  {author} {\bibinfo {author} {\bibfnamefont {S.~L.}\ \bibnamefont
  {Detweiler}},\ }\href {\doibase 10.1103/PhysRevD.22.2323} {\bibfield
  {journal} {\bibinfo  {journal} {Phys. Rev.}\ }\textbf {\bibinfo {volume}
  {D22}},\ \bibinfo {pages} {2323} (\bibinfo {year} {1980})}\BibitemShut
  {NoStop}%
\bibitem [{\citenamefont {Cardoso}\ \emph {et~al.}(2004)\citenamefont
  {Cardoso}, \citenamefont {Dias}, \citenamefont {Lemos},\ and\ \citenamefont
  {Yoshida}}]{Cardoso:2004nk}%
  \BibitemOpen
  \bibfield  {author} {\bibinfo {author} {\bibfnamefont {V.}~\bibnamefont
  {Cardoso}}, \bibinfo {author} {\bibfnamefont {O.~J.}\ \bibnamefont {Dias}},
  \bibinfo {author} {\bibfnamefont {J.~P.}\ \bibnamefont {Lemos}}, \ and\
  \bibinfo {author} {\bibfnamefont {S.}~\bibnamefont {Yoshida}},\ }\href
  {\doibase 10.1103/PhysRevD.70.049903, 10.1103/PhysRevD.70.044039} {\bibfield
  {journal} {\bibinfo  {journal} {Phys.Rev.}\ }\textbf {\bibinfo {volume}
  {D70}},\ \bibinfo {pages} {044039} (\bibinfo {year} {2004})},\ \Eprint
  {http://arxiv.org/abs/hep-th/0404096} {arXiv:hep-th/0404096 [hep-th]}
  \BibitemShut {NoStop}%
\bibitem [{\citenamefont {Brito}\ \emph
  {et~al.}(2015{\natexlab{a}})\citenamefont {Brito}, \citenamefont {Cardoso},\
  and\ \citenamefont {Pani}}]{Brito:2015oca}%
  \BibitemOpen
  \bibfield  {author} {\bibinfo {author} {\bibfnamefont {R.}~\bibnamefont
  {Brito}}, \bibinfo {author} {\bibfnamefont {V.}~\bibnamefont {Cardoso}}, \
  and\ \bibinfo {author} {\bibfnamefont {P.}~\bibnamefont {Pani}},\ }\href
  {\doibase 10.1007/978-3-319-19000-6} {\bibfield  {journal} {\bibinfo
  {journal} {Lect. Notes Phys.}\ }\textbf {\bibinfo {volume} {906}},\ \bibinfo
  {pages} {pp.1} (\bibinfo {year} {2015}{\natexlab{a}})},\ \Eprint
  {http://arxiv.org/abs/1501.06570} {arXiv:1501.06570 [gr-qc]} \BibitemShut
  {NoStop}%
\bibitem [{\citenamefont
  {Shlapentokh-Rothman}(2014)}]{Shlapentokh-Rothman:2013ysa}%
  \BibitemOpen
  \bibfield  {author} {\bibinfo {author} {\bibfnamefont {Y.}~\bibnamefont
  {Shlapentokh-Rothman}},\ }\href {\doibase 10.1007/s00220-014-2033-x}
  {\bibfield  {journal} {\bibinfo  {journal} {Commun.Math.Phys.}\ }\textbf
  {\bibinfo {volume} {329}},\ \bibinfo {pages} {859} (\bibinfo {year}
  {2014})},\ \Eprint {http://arxiv.org/abs/1302.3448} {arXiv:1302.3448 [gr-qc]}
  \BibitemShut {NoStop}%
\bibitem [{\citenamefont {Damour}\ \emph {et~al.}(1976)\citenamefont {Damour},
  \citenamefont {Deruelle},\ and\ \citenamefont {Ruffini}}]{Damour:1976kh}%
  \BibitemOpen
  \bibfield  {author} {\bibinfo {author} {\bibfnamefont {T.}~\bibnamefont
  {Damour}}, \bibinfo {author} {\bibfnamefont {N.}~\bibnamefont {Deruelle}}, \
  and\ \bibinfo {author} {\bibfnamefont {R.}~\bibnamefont {Ruffini}},\ }\href
  {\doibase 10.1007/BF02725534} {\bibfield  {journal} {\bibinfo  {journal}
  {Lett.Nuovo Cim.}\ }\textbf {\bibinfo {volume} {15}},\ \bibinfo {pages} {257}
  (\bibinfo {year} {1976})}\BibitemShut {NoStop}%
\bibitem [{\citenamefont {Zouros}\ and\ \citenamefont
  {Eardley}(1979)}]{Zouros:1979iw}%
  \BibitemOpen
  \bibfield  {author} {\bibinfo {author} {\bibfnamefont {T.}~\bibnamefont
  {Zouros}}\ and\ \bibinfo {author} {\bibfnamefont {D.}~\bibnamefont
  {Eardley}},\ }\href {\doibase 10.1016/0003-4916(79)90237-9} {\bibfield
  {journal} {\bibinfo  {journal} {Annals Phys.}\ }\textbf {\bibinfo {volume}
  {118}},\ \bibinfo {pages} {139} (\bibinfo {year} {1979})}\BibitemShut
  {NoStop}%
\bibitem [{\citenamefont {Dolan}(2007)}]{Dolan:2007mj}%
  \BibitemOpen
  \bibfield  {author} {\bibinfo {author} {\bibfnamefont {S.~R.}\ \bibnamefont
  {Dolan}},\ }\href {\doibase 10.1103/PhysRevD.76.084001} {\bibfield  {journal}
  {\bibinfo  {journal} {Phys.Rev.}\ }\textbf {\bibinfo {volume} {D76}},\
  \bibinfo {pages} {084001} (\bibinfo {year} {2007})},\ \Eprint
  {http://arxiv.org/abs/0705.2880} {arXiv:0705.2880 [gr-qc]} \BibitemShut
  {NoStop}%
\bibitem [{\citenamefont {Pani}\ \emph
  {et~al.}(2012{\natexlab{a}})\citenamefont {Pani}, \citenamefont {Cardoso},
  \citenamefont {Gualtieri}, \citenamefont {Berti},\ and\ \citenamefont
  {Ishibashi}}]{Pani:2012vp}%
  \BibitemOpen
  \bibfield  {author} {\bibinfo {author} {\bibfnamefont {P.}~\bibnamefont
  {Pani}}, \bibinfo {author} {\bibfnamefont {V.}~\bibnamefont {Cardoso}},
  \bibinfo {author} {\bibfnamefont {L.}~\bibnamefont {Gualtieri}}, \bibinfo
  {author} {\bibfnamefont {E.}~\bibnamefont {Berti}}, \ and\ \bibinfo {author}
  {\bibfnamefont {A.}~\bibnamefont {Ishibashi}},\ }\href {\doibase
  10.1103/PhysRevLett.109.131102} {\bibfield  {journal} {\bibinfo  {journal}
  {Phys.Rev.Lett.}\ }\textbf {\bibinfo {volume} {109}},\ \bibinfo {pages}
  {131102} (\bibinfo {year} {2012}{\natexlab{a}})},\ \Eprint
  {http://arxiv.org/abs/1209.0465} {arXiv:1209.0465 [gr-qc]} \BibitemShut
  {NoStop}%
\bibitem [{\citenamefont {Pani}\ \emph
  {et~al.}(2012{\natexlab{b}})\citenamefont {Pani}, \citenamefont {Cardoso},
  \citenamefont {Gualtieri}, \citenamefont {Berti},\ and\ \citenamefont
  {Ishibashi}}]{Pani:2012bp}%
  \BibitemOpen
  \bibfield  {author} {\bibinfo {author} {\bibfnamefont {P.}~\bibnamefont
  {Pani}}, \bibinfo {author} {\bibfnamefont {V.}~\bibnamefont {Cardoso}},
  \bibinfo {author} {\bibfnamefont {L.}~\bibnamefont {Gualtieri}}, \bibinfo
  {author} {\bibfnamefont {E.}~\bibnamefont {Berti}}, \ and\ \bibinfo {author}
  {\bibfnamefont {A.}~\bibnamefont {Ishibashi}},\ }\href {\doibase
  10.1103/PhysRevD.86.104017} {\bibfield  {journal} {\bibinfo  {journal}
  {Phys.Rev.}\ }\textbf {\bibinfo {volume} {D86}},\ \bibinfo {pages} {104017}
  (\bibinfo {year} {2012}{\natexlab{b}})},\ \Eprint
  {http://arxiv.org/abs/1209.0773} {arXiv:1209.0773 [gr-qc]} \BibitemShut
  {NoStop}%
\bibitem [{\citenamefont {Witek}\ \emph {et~al.}(2013)\citenamefont {Witek},
  \citenamefont {Cardoso}, \citenamefont {Ishibashi},\ and\ \citenamefont
  {Sperhake}}]{Witek:2012tr}%
  \BibitemOpen
  \bibfield  {author} {\bibinfo {author} {\bibfnamefont {H.}~\bibnamefont
  {Witek}}, \bibinfo {author} {\bibfnamefont {V.}~\bibnamefont {Cardoso}},
  \bibinfo {author} {\bibfnamefont {A.}~\bibnamefont {Ishibashi}}, \ and\
  \bibinfo {author} {\bibfnamefont {U.}~\bibnamefont {Sperhake}},\ }\href
  {\doibase 10.1103/PhysRevD.87.043513} {\bibfield  {journal} {\bibinfo
  {journal} {Phys.Rev.}\ }\textbf {\bibinfo {volume} {D87}},\ \bibinfo {pages}
  {043513} (\bibinfo {year} {2013})},\ \Eprint {http://arxiv.org/abs/1212.0551}
  {arXiv:1212.0551 [gr-qc]} \BibitemShut {NoStop}%
\bibitem [{\citenamefont {Endlich}\ and\ \citenamefont
  {Penco}(2017)}]{Endlich:2016jgc}%
  \BibitemOpen
  \bibfield  {author} {\bibinfo {author} {\bibfnamefont {S.}~\bibnamefont
  {Endlich}}\ and\ \bibinfo {author} {\bibfnamefont {R.}~\bibnamefont
  {Penco}},\ }\href {\doibase 10.1007/JHEP05(2017)052} {\bibfield  {journal}
  {\bibinfo  {journal} {JHEP}\ }\textbf {\bibinfo {volume} {05}},\ \bibinfo
  {pages} {052} (\bibinfo {year} {2017})},\ \Eprint
  {http://arxiv.org/abs/1609.06723} {arXiv:1609.06723 [hep-th]} \BibitemShut
  {NoStop}%
\bibitem [{\citenamefont {East}(2017)}]{East:2017mrj}%
  \BibitemOpen
  \bibfield  {author} {\bibinfo {author} {\bibfnamefont {W.~E.}\ \bibnamefont
  {East}},\ }\href {\doibase 10.1103/PhysRevD.96.024004} {\bibfield  {journal}
  {\bibinfo  {journal} {Phys. Rev.}\ }\textbf {\bibinfo {volume} {D96}},\
  \bibinfo {pages} {024004} (\bibinfo {year} {2017})},\ \Eprint
  {http://arxiv.org/abs/1705.01544} {arXiv:1705.01544 [gr-qc]} \BibitemShut
  {NoStop}%
\bibitem [{\citenamefont {East}\ and\ \citenamefont
  {Pretorius}(2017)}]{East:2017ovw}%
  \BibitemOpen
  \bibfield  {author} {\bibinfo {author} {\bibfnamefont {W.~E.}\ \bibnamefont
  {East}}\ and\ \bibinfo {author} {\bibfnamefont {F.}~\bibnamefont
  {Pretorius}},\ }\href {\doibase 10.1103/PhysRevLett.119.041101} {\bibfield
  {journal} {\bibinfo  {journal} {Phys. Rev. Lett.}\ }\textbf {\bibinfo
  {volume} {119}},\ \bibinfo {pages} {041101} (\bibinfo {year} {2017})},\
  \Eprint {http://arxiv.org/abs/1704.04791} {arXiv:1704.04791 [gr-qc]}
  \BibitemShut {NoStop}%
\bibitem [{\citenamefont {Baryakhtar}\ \emph {et~al.}(2017)\citenamefont
  {Baryakhtar}, \citenamefont {Lasenby},\ and\ \citenamefont
  {Teo}}]{Baryakhtar:2017ngi}%
  \BibitemOpen
  \bibfield  {author} {\bibinfo {author} {\bibfnamefont {M.}~\bibnamefont
  {Baryakhtar}}, \bibinfo {author} {\bibfnamefont {R.}~\bibnamefont {Lasenby}},
  \ and\ \bibinfo {author} {\bibfnamefont {M.}~\bibnamefont {Teo}},\ }\href
  {\doibase 10.1103/PhysRevD.96.035019} {\bibfield  {journal} {\bibinfo
  {journal} {Phys. Rev.}\ }\textbf {\bibinfo {volume} {D96}},\ \bibinfo {pages}
  {035019} (\bibinfo {year} {2017})},\ \Eprint
  {http://arxiv.org/abs/1704.05081} {arXiv:1704.05081 [hep-ph]} \BibitemShut
  {NoStop}%
\bibitem [{\citenamefont {East}(2018)}]{East:2018glu}%
  \BibitemOpen
  \bibfield  {author} {\bibinfo {author} {\bibfnamefont {W.~E.}\ \bibnamefont
  {East}},\ }\href {\doibase 10.1103/PhysRevLett.121.131104} {\bibfield
  {journal} {\bibinfo  {journal} {Phys. Rev. Lett.}\ }\textbf {\bibinfo
  {volume} {121}},\ \bibinfo {pages} {131104} (\bibinfo {year} {2018})},\
  \Eprint {http://arxiv.org/abs/1807.00043} {arXiv:1807.00043 [gr-qc]}
  \BibitemShut {NoStop}%
\bibitem [{\citenamefont {Frolov}\ \emph {et~al.}(2018)\citenamefont {Frolov},
  \citenamefont {Krtous}, \citenamefont {Kubiznak},\ and\ \citenamefont
  {Santos}}]{Frolov:2018ezx}%
  \BibitemOpen
  \bibfield  {author} {\bibinfo {author} {\bibfnamefont {V.~P.}\ \bibnamefont
  {Frolov}}, \bibinfo {author} {\bibfnamefont {P.}~\bibnamefont {Krtous}},
  \bibinfo {author} {\bibfnamefont {D.}~\bibnamefont {Kubiznak}}, \ and\
  \bibinfo {author} {\bibfnamefont {J.~E.}\ \bibnamefont {Santos}},\ }\href
  {\doibase 10.1103/PhysRevLett.120.231103} {\bibfield  {journal} {\bibinfo
  {journal} {Phys. Rev. Lett.}\ }\textbf {\bibinfo {volume} {120}},\ \bibinfo
  {pages} {231103} (\bibinfo {year} {2018})},\ \Eprint
  {http://arxiv.org/abs/1804.00030} {arXiv:1804.00030 [hep-th]} \BibitemShut
  {NoStop}%
\bibitem [{\citenamefont {Dolan}(2018)}]{Dolan:2018dqv}%
  \BibitemOpen
  \bibfield  {author} {\bibinfo {author} {\bibfnamefont {S.~R.}\ \bibnamefont
  {Dolan}},\ }\href {\doibase 10.1103/PhysRevD.98.104006} {\bibfield  {journal}
  {\bibinfo  {journal} {Phys. Rev.}\ }\textbf {\bibinfo {volume} {D98}},\
  \bibinfo {pages} {104006} (\bibinfo {year} {2018})},\ \Eprint
  {http://arxiv.org/abs/1806.01604} {arXiv:1806.01604 [gr-qc]} \BibitemShut
  {NoStop}%
\bibitem [{\citenamefont {Siemonsen}\ and\ \citenamefont
  {East}(2020)}]{Siemonsen:2019ebd}%
  \BibitemOpen
  \bibfield  {author} {\bibinfo {author} {\bibfnamefont {N.}~\bibnamefont
  {Siemonsen}}\ and\ \bibinfo {author} {\bibfnamefont {W.~E.}\ \bibnamefont
  {East}},\ }\href {\doibase 10.1103/PhysRevD.101.024019} {\bibfield  {journal}
  {\bibinfo  {journal} {Phys. Rev.}\ }\textbf {\bibinfo {volume} {D101}},\
  \bibinfo {pages} {024019} (\bibinfo {year} {2020})},\ \Eprint
  {http://arxiv.org/abs/1910.09476} {arXiv:1910.09476 [gr-qc]} \BibitemShut
  {NoStop}%
\bibitem [{\citenamefont {Brito}\ \emph
  {et~al.}(2013{\natexlab{a}})\citenamefont {Brito}, \citenamefont {Cardoso},\
  and\ \citenamefont {Pani}}]{Brito:2013wya}%
  \BibitemOpen
  \bibfield  {author} {\bibinfo {author} {\bibfnamefont {R.}~\bibnamefont
  {Brito}}, \bibinfo {author} {\bibfnamefont {V.}~\bibnamefont {Cardoso}}, \
  and\ \bibinfo {author} {\bibfnamefont {P.}~\bibnamefont {Pani}},\ }\href
  {\doibase 10.1103/PhysRevD.88.023514} {\bibfield  {journal} {\bibinfo
  {journal} {Phys. Rev.}\ }\textbf {\bibinfo {volume} {D88}},\ \bibinfo {pages}
  {023514} (\bibinfo {year} {2013}{\natexlab{a}})},\ \Eprint
  {http://arxiv.org/abs/1304.6725} {arXiv:1304.6725 [gr-qc]} \BibitemShut
  {NoStop}%
\bibitem [{\citenamefont {Hinterbichler}(2012)}]{Hinterbichler:2011tt}%
  \BibitemOpen
  \bibfield  {author} {\bibinfo {author} {\bibfnamefont {K.}~\bibnamefont
  {Hinterbichler}},\ }\href {\doibase 10.1103/RevModPhys.84.671} {\bibfield
  {journal} {\bibinfo  {journal} {Rev.Mod.Phys.}\ }\textbf {\bibinfo {volume}
  {84}},\ \bibinfo {pages} {671} (\bibinfo {year} {2012})},\ \Eprint
  {http://arxiv.org/abs/1105.3735} {arXiv:1105.3735 [hep-th]} \BibitemShut
  {NoStop}%
\bibitem [{\citenamefont {de~Rham}\ \emph {et~al.}(2011)\citenamefont
  {de~Rham}, \citenamefont {Gabadadze},\ and\ \citenamefont
  {Tolley}}]{deRham:2010kj}%
  \BibitemOpen
  \bibfield  {author} {\bibinfo {author} {\bibfnamefont {C.}~\bibnamefont
  {de~Rham}}, \bibinfo {author} {\bibfnamefont {G.}~\bibnamefont {Gabadadze}},
  \ and\ \bibinfo {author} {\bibfnamefont {A.~J.}\ \bibnamefont {Tolley}},\
  }\href {\doibase 10.1103/PhysRevLett.106.231101} {\bibfield  {journal}
  {\bibinfo  {journal} {Phys. Rev. Lett.}\ }\textbf {\bibinfo {volume} {106}},\
  \bibinfo {pages} {231101} (\bibinfo {year} {2011})},\ \Eprint
  {http://arxiv.org/abs/1011.1232} {arXiv:1011.1232 [hep-th]} \BibitemShut
  {NoStop}%
\bibitem [{\citenamefont {Hassan}\ and\ \citenamefont
  {Rosen}(2012{\natexlab{a}})}]{Hassan:2011hr}%
  \BibitemOpen
  \bibfield  {author} {\bibinfo {author} {\bibfnamefont {S.~F.}\ \bibnamefont
  {Hassan}}\ and\ \bibinfo {author} {\bibfnamefont {R.~A.}\ \bibnamefont
  {Rosen}},\ }\href {\doibase 10.1103/PhysRevLett.108.041101} {\bibfield
  {journal} {\bibinfo  {journal} {Phys. Rev. Lett.}\ }\textbf {\bibinfo
  {volume} {108}},\ \bibinfo {pages} {041101} (\bibinfo {year}
  {2012}{\natexlab{a}})},\ \Eprint {http://arxiv.org/abs/1106.3344}
  {arXiv:1106.3344 [hep-th]} \BibitemShut {NoStop}%
\bibitem [{\citenamefont {Hassan}\ and\ \citenamefont
  {Rosen}(2012{\natexlab{b}})}]{Hassan:2011zd}%
  \BibitemOpen
  \bibfield  {author} {\bibinfo {author} {\bibfnamefont {S.~F.}\ \bibnamefont
  {Hassan}}\ and\ \bibinfo {author} {\bibfnamefont {R.~A.}\ \bibnamefont
  {Rosen}},\ }\href {\doibase 10.1007/JHEP02(2012)126} {\bibfield  {journal}
  {\bibinfo  {journal} {JHEP}\ }\textbf {\bibinfo {volume} {02}},\ \bibinfo
  {pages} {126} (\bibinfo {year} {2012}{\natexlab{b}})},\ \Eprint
  {http://arxiv.org/abs/1109.3515} {arXiv:1109.3515 [hep-th]} \BibitemShut
  {NoStop}%
\bibitem [{\citenamefont {de~Rham}(2014)}]{deRham:2014zqa}%
  \BibitemOpen
  \bibfield  {author} {\bibinfo {author} {\bibfnamefont {C.}~\bibnamefont
  {de~Rham}},\ }\href {\doibase 10.12942/lrr-2014-7} {\bibfield  {journal}
  {\bibinfo  {journal} {Living Rev.Rel.}\ }\textbf {\bibinfo {volume} {17}},\
  \bibinfo {pages} {7} (\bibinfo {year} {2014})},\ \Eprint
  {http://arxiv.org/abs/1401.4173} {arXiv:1401.4173 [hep-th]} \BibitemShut
  {NoStop}%
\bibitem [{deS()}]{deSitter}%
  \BibitemOpen
  \href@noop {} {}\bibinfo {note} {The extension to Einstein spacetimes is
  straighforward; we assume zero cosmological constant for
  simplicity.}\BibitemShut {Stop}%
\bibitem [{\citenamefont {Babichev}\ \emph {et~al.}(2016)\citenamefont
  {Babichev}, \citenamefont {Marzola}, \citenamefont {Raidal}, \citenamefont
  {Schmidt-May}, \citenamefont {Urban}, \citenamefont {Veermäe},\ and\
  \citenamefont {von Strauss}}]{Babichev:2016bxi}%
  \BibitemOpen
  \bibfield  {author} {\bibinfo {author} {\bibfnamefont {E.}~\bibnamefont
  {Babichev}}, \bibinfo {author} {\bibfnamefont {L.}~\bibnamefont {Marzola}},
  \bibinfo {author} {\bibfnamefont {M.}~\bibnamefont {Raidal}}, \bibinfo
  {author} {\bibfnamefont {A.}~\bibnamefont {Schmidt-May}}, \bibinfo {author}
  {\bibfnamefont {F.}~\bibnamefont {Urban}}, \bibinfo {author} {\bibfnamefont
  {H.}~\bibnamefont {Veermäe}}, \ and\ \bibinfo {author} {\bibfnamefont
  {M.}~\bibnamefont {von Strauss}},\ }\href {\doibase
  10.1088/1475-7516/2016/09/016} {\bibfield  {journal} {\bibinfo  {journal}
  {JCAP}\ }\textbf {\bibinfo {volume} {1609}},\ \bibinfo {pages} {016}
  (\bibinfo {year} {2016})},\ \Eprint {http://arxiv.org/abs/1607.03497}
  {arXiv:1607.03497 [hep-th]} \BibitemShut {NoStop}%
\bibitem [{\citenamefont {Brito}\ \emph
  {et~al.}(2013{\natexlab{b}})\citenamefont {Brito}, \citenamefont {Cardoso},\
  and\ \citenamefont {Pani}}]{Brito:2013xaa}%
  \BibitemOpen
  \bibfield  {author} {\bibinfo {author} {\bibfnamefont {R.}~\bibnamefont
  {Brito}}, \bibinfo {author} {\bibfnamefont {V.}~\bibnamefont {Cardoso}}, \
  and\ \bibinfo {author} {\bibfnamefont {P.}~\bibnamefont {Pani}},\ }\href
  {\doibase 10.1103/PhysRevD.88.064006} {\bibfield  {journal} {\bibinfo
  {journal} {Phys. Rev.}\ }\textbf {\bibinfo {volume} {D88}},\ \bibinfo {pages}
  {064006} (\bibinfo {year} {2013}{\natexlab{b}})},\ \Eprint
  {http://arxiv.org/abs/1309.0818} {arXiv:1309.0818 [gr-qc]} \BibitemShut
  {NoStop}%
\bibitem [{\citenamefont {Babichev}\ and\ \citenamefont
  {Brito}(2015)}]{Babichev:2015xha}%
  \BibitemOpen
  \bibfield  {author} {\bibinfo {author} {\bibfnamefont {E.}~\bibnamefont
  {Babichev}}\ and\ \bibinfo {author} {\bibfnamefont {R.}~\bibnamefont
  {Brito}},\ }\href {\doibase 10.1088/0264-9381/32/15/154001} {\bibfield
  {journal} {\bibinfo  {journal} {Class. Quant. Grav.}\ }\textbf {\bibinfo
  {volume} {32}},\ \bibinfo {pages} {154001} (\bibinfo {year} {2015})},\
  \Eprint {http://arxiv.org/abs/1503.07529} {arXiv:1503.07529 [gr-qc]}
  \BibitemShut {NoStop}%
\bibitem [{\citenamefont {Mazuet}\ and\ \citenamefont
  {Volkov}(2018)}]{Mazuet:2018ysa}%
  \BibitemOpen
  \bibfield  {author} {\bibinfo {author} {\bibfnamefont {C.}~\bibnamefont
  {Mazuet}}\ and\ \bibinfo {author} {\bibfnamefont {M.~S.}\ \bibnamefont
  {Volkov}},\ }\href {\doibase 10.1088/1475-7516/2018/07/012} {\bibfield
  {journal} {\bibinfo  {journal} {JCAP}\ }\textbf {\bibinfo {volume} {1807}},\
  \bibinfo {pages} {012} (\bibinfo {year} {2018})},\ \Eprint
  {http://arxiv.org/abs/1804.01970} {arXiv:1804.01970 [hep-th]} \BibitemShut
  {NoStop}%
\bibitem [{\citenamefont {Babichev}\ and\ \citenamefont
  {Fabbri}(2013)}]{Babichev:2013una}%
  \BibitemOpen
  \bibfield  {author} {\bibinfo {author} {\bibfnamefont {E.}~\bibnamefont
  {Babichev}}\ and\ \bibinfo {author} {\bibfnamefont {A.}~\bibnamefont
  {Fabbri}},\ }\href {\doibase 10.1088/0264-9381/30/15/152001} {\bibfield
  {journal} {\bibinfo  {journal} {Class.Quant.Grav.}\ }\textbf {\bibinfo
  {volume} {30}},\ \bibinfo {pages} {152001} (\bibinfo {year} {2013})},\
  \Eprint {http://arxiv.org/abs/1304.5992} {arXiv:1304.5992 [gr-qc]}
  \BibitemShut {NoStop}%
\bibitem [{\citenamefont {Mathews}(1962)}]{Mathews:1962}%
  \BibitemOpen
  \bibfield  {author} {\bibinfo {author} {\bibfnamefont {J.}~\bibnamefont
  {Mathews}},\ }\href {http://www.jstor.org/stable/2098922} {\bibfield
  {journal} {\bibinfo  {journal} {Journal of the Society for Industrial and
  Applied Mathematics}\ }\textbf {\bibinfo {volume} {10}},\ \bibinfo {pages}
  {768} (\bibinfo {year} {1962})}\BibitemShut {NoStop}%
\bibitem [{\citenamefont {Thorne}(1980)}]{Thorne:1980ru}%
  \BibitemOpen
  \bibfield  {author} {\bibinfo {author} {\bibfnamefont {K.~S.}\ \bibnamefont
  {Thorne}},\ }\href {\doibase 10.1103/RevModPhys.52.299} {\bibfield  {journal}
  {\bibinfo  {journal} {Rev. Mod. Phys.}\ }\textbf {\bibinfo {volume} {52}},\
  \bibinfo {pages} {299} (\bibinfo {year} {1980})}\BibitemShut {NoStop}%
\bibitem [{\citenamefont {Maggiore}(2007)}]{Maggiore:1900zz}%
  \BibitemOpen
  \bibfield  {author} {\bibinfo {author} {\bibfnamefont {M.}~\bibnamefont
  {Maggiore}},\ }\href {http://www.oup.com/uk/catalogue/?ci=9780198570745}
  {\emph {\bibinfo {title} {{Gravitational Waves. Vol. 1: Theory and
  Experiments}}}},\ Oxford Master Series in Physics\ (\bibinfo  {publisher}
  {Oxford University Press},\ \bibinfo {year} {2007})\BibitemShut {NoStop}%
\bibitem [{\citenamefont {Aoki}\ \emph {et~al.}(2018)\citenamefont {Aoki},
  \citenamefont {Maeda}, \citenamefont {Misonoh},\ and\ \citenamefont
  {Okawa}}]{Aoki:2017ixz}%
  \BibitemOpen
  \bibfield  {author} {\bibinfo {author} {\bibfnamefont {K.}~\bibnamefont
  {Aoki}}, \bibinfo {author} {\bibfnamefont {K.-i.}\ \bibnamefont {Maeda}},
  \bibinfo {author} {\bibfnamefont {Y.}~\bibnamefont {Misonoh}}, \ and\
  \bibinfo {author} {\bibfnamefont {H.}~\bibnamefont {Okawa}},\ }\href
  {\doibase 10.1103/PhysRevD.97.044005} {\bibfield  {journal} {\bibinfo
  {journal} {Phys. Rev.}\ }\textbf {\bibinfo {volume} {D97}},\ \bibinfo {pages}
  {044005} (\bibinfo {year} {2018})},\ \Eprint
  {http://arxiv.org/abs/1710.05606} {arXiv:1710.05606 [gr-qc]} \BibitemShut
  {NoStop}%
\bibitem [{\citenamefont {{Starobinskij}}(1973)}]{1973ZhETF..64...48S}%
  \BibitemOpen
  \bibfield  {author} {\bibinfo {author} {\bibfnamefont {A.~A.}\ \bibnamefont
  {{Starobinskij}}},\ }\href@noop {} {\bibfield  {journal} {\bibinfo  {journal}
  {Zhurnal Eksperimentalnoi i Teoreticheskoi Fiziki}\ }\textbf {\bibinfo
  {volume} {64}},\ \bibinfo {pages} {48} (\bibinfo {year} {1973})}\BibitemShut
  {NoStop}%
\bibitem [{\citenamefont {{Starobinskij}}\ and\ \citenamefont
  {{Churilov}}(1973)}]{1973ZhETF..65....3S}%
  \BibitemOpen
  \bibfield  {author} {\bibinfo {author} {\bibfnamefont {A.~A.}\ \bibnamefont
  {{Starobinskij}}}\ and\ \bibinfo {author} {\bibfnamefont {S.~M.}\
  \bibnamefont {{Churilov}}},\ }\href@noop {} {\bibfield  {journal} {\bibinfo
  {journal} {Zhurnal Eksperimentalnoi i Teoreticheskoi Fiziki}\ }\textbf
  {\bibinfo {volume} {65}},\ \bibinfo {pages} {3} (\bibinfo {year}
  {1973})}\BibitemShut {NoStop}%
\bibitem [{\citenamefont {Brito}\ \emph
  {et~al.}(2015{\natexlab{b}})\citenamefont {Brito}, \citenamefont {Cardoso},\
  and\ \citenamefont {Pani}}]{Brito:2014wla}%
  \BibitemOpen
  \bibfield  {author} {\bibinfo {author} {\bibfnamefont {R.}~\bibnamefont
  {Brito}}, \bibinfo {author} {\bibfnamefont {V.}~\bibnamefont {Cardoso}}, \
  and\ \bibinfo {author} {\bibfnamefont {P.}~\bibnamefont {Pani}},\ }\href
  {\doibase 10.1088/0264-9381/32/13/134001} {\bibfield  {journal} {\bibinfo
  {journal} {Class. Quant. Grav.}\ }\textbf {\bibinfo {volume} {32}},\ \bibinfo
  {pages} {134001} (\bibinfo {year} {2015}{\natexlab{b}})},\ \Eprint
  {http://arxiv.org/abs/1411.0686} {arXiv:1411.0686 [gr-qc]} \BibitemShut
  {NoStop}%
\bibitem [{\citenamefont {Teukolsky}(1973)}]{Teukolsky:1973ha}%
  \BibitemOpen
  \bibfield  {author} {\bibinfo {author} {\bibfnamefont {S.~A.}\ \bibnamefont
  {Teukolsky}},\ }\href {\doibase 10.1086/152444} {\bibfield  {journal}
  {\bibinfo  {journal} {Astrophys. J.}\ }\textbf {\bibinfo {volume} {185}},\
  \bibinfo {pages} {635} (\bibinfo {year} {1973})}\BibitemShut {NoStop}%
\bibitem [{\citenamefont {Yoshino}\ and\ \citenamefont
  {Kodama}(2014)}]{Yoshino:2013ofa}%
  \BibitemOpen
  \bibfield  {author} {\bibinfo {author} {\bibfnamefont {H.}~\bibnamefont
  {Yoshino}}\ and\ \bibinfo {author} {\bibfnamefont {H.}~\bibnamefont
  {Kodama}},\ }\href {\doibase 10.1093/ptep/ptu029} {\bibfield  {journal}
  {\bibinfo  {journal} {PTEP}\ }\textbf {\bibinfo {volume} {2014}},\ \bibinfo
  {pages} {043E02} (\bibinfo {year} {2014})},\ \Eprint
  {http://arxiv.org/abs/1312.2326} {arXiv:1312.2326 [gr-qc]} \BibitemShut
  {NoStop}%
\bibitem [{\citenamefont {Poisson}(1993)}]{Poisson:1993vp}%
  \BibitemOpen
  \bibfield  {author} {\bibinfo {author} {\bibfnamefont {E.}~\bibnamefont
  {Poisson}},\ }\href {\doibase 10.1103/PhysRevD.47.1497} {\bibfield  {journal}
  {\bibinfo  {journal} {Phys. Rev.}\ }\textbf {\bibinfo {volume} {D47}},\
  \bibinfo {pages} {1497} (\bibinfo {year} {1993})}\BibitemShut {NoStop}%
\bibitem [{\citenamefont {Ficarra}\ \emph {et~al.}(2019)\citenamefont
  {Ficarra}, \citenamefont {Pani},\ and\ \citenamefont
  {Witek}}]{Ficarra:2018rfu}%
  \BibitemOpen
  \bibfield  {author} {\bibinfo {author} {\bibfnamefont {G.}~\bibnamefont
  {Ficarra}}, \bibinfo {author} {\bibfnamefont {P.}~\bibnamefont {Pani}}, \
  and\ \bibinfo {author} {\bibfnamefont {H.}~\bibnamefont {Witek}},\ }\href
  {\doibase 10.1103/PhysRevD.99.104019} {\bibfield  {journal} {\bibinfo
  {journal} {Phys. Rev.}\ }\textbf {\bibinfo {volume} {D99}},\ \bibinfo {pages}
  {104019} (\bibinfo {year} {2019})},\ \Eprint
  {http://arxiv.org/abs/1812.02758} {arXiv:1812.02758 [gr-qc]} \BibitemShut
  {NoStop}%
\bibitem [{\citenamefont {Abbott}\ \emph {et~al.}(2013)\citenamefont {Abbott}
  \emph {et~al.}}]{Aasi:2013wya}%
  \BibitemOpen
  \bibfield  {author} {\bibinfo {author} {\bibfnamefont {B.~P.}\ \bibnamefont
  {Abbott}} \emph {et~al.} (\bibinfo {collaboration} {VIRGO, LIGO
  Scientific}),\ }\href {\doibase 10.1007/lrr-2016-1} {\  (\bibinfo {year}
  {2013}),\ 10.1007/lrr-2016-1},\ \bibinfo {note} {[Living Rev.
  Rel.19,1(2016)]},\ \Eprint {http://arxiv.org/abs/1304.0670} {arXiv:1304.0670
  [gr-qc]} \BibitemShut {NoStop}%
\bibitem [{\citenamefont {Robson}\ \emph {et~al.}(2019)\citenamefont {Robson},
  \citenamefont {Cornish},\ and\ \citenamefont {Liug}}]{Cornish:2018dyw}%
  \BibitemOpen
  \bibfield  {author} {\bibinfo {author} {\bibfnamefont {T.}~\bibnamefont
  {Robson}}, \bibinfo {author} {\bibfnamefont {N.~J.}\ \bibnamefont {Cornish}},
  \ and\ \bibinfo {author} {\bibfnamefont {C.}~\bibnamefont {Liug}},\ }\href
  {\doibase 10.1088/1361-6382/ab1101} {\bibfield  {journal} {\bibinfo
  {journal} {Class. Quant. Grav.}\ }\textbf {\bibinfo {volume} {36}},\ \bibinfo
  {pages} {105011} (\bibinfo {year} {2019})},\ \Eprint
  {http://arxiv.org/abs/1803.01944} {arXiv:1803.01944 [astro-ph.HE]}
  \BibitemShut {NoStop}%
\bibitem [{\citenamefont {Kawamura}\ \emph {et~al.}(2006)\citenamefont
  {Kawamura} \emph {et~al.}}]{Kawamura:2006up}%
  \BibitemOpen
  \bibfield  {author} {\bibinfo {author} {\bibfnamefont {S.}~\bibnamefont
  {Kawamura}} \emph {et~al.},\ }\bibfield  {booktitle} {\emph {\bibinfo
  {booktitle} {{Gravitational waves. Proceedings, 6th Edoardo Amaldi
  Conference, Amaldi6, Bankoku Shinryoukan, June 20-24, 2005}}},\ }\href
  {\doibase 10.1088/0264-9381/23/8/S17} {\bibfield  {journal} {\bibinfo
  {journal} {Class. Quant. Grav.}\ }\textbf {\bibinfo {volume} {23}},\ \bibinfo
  {pages} {S125} (\bibinfo {year} {2006})}\BibitemShut {NoStop}%
\bibitem [{\citenamefont {Arvanitaki}\ and\ \citenamefont
  {Dubovsky}(2011)}]{Arvanitaki:2010sy}%
  \BibitemOpen
  \bibfield  {author} {\bibinfo {author} {\bibfnamefont {A.}~\bibnamefont
  {Arvanitaki}}\ and\ \bibinfo {author} {\bibfnamefont {S.}~\bibnamefont
  {Dubovsky}},\ }\href {\doibase 10.1103/PhysRevD.83.044026} {\bibfield
  {journal} {\bibinfo  {journal} {Phys. Rev.}\ }\textbf {\bibinfo {volume}
  {D83}},\ \bibinfo {pages} {044026} (\bibinfo {year} {2011})},\ \Eprint
  {http://arxiv.org/abs/1004.3558} {arXiv:1004.3558 [hep-th]} \BibitemShut
  {NoStop}%
\bibitem [{\citenamefont {Ng}\ \emph {et~al.}(2019)\citenamefont {Ng},
  \citenamefont {Hannuksela}, \citenamefont {Vitale},\ and\ \citenamefont
  {Li}}]{Ng:2019jsx}%
  \BibitemOpen
  \bibfield  {author} {\bibinfo {author} {\bibfnamefont {K.~K.~Y.}\
  \bibnamefont {Ng}}, \bibinfo {author} {\bibfnamefont {O.~A.}\ \bibnamefont
  {Hannuksela}}, \bibinfo {author} {\bibfnamefont {S.}~\bibnamefont {Vitale}},
  \ and\ \bibinfo {author} {\bibfnamefont {T.~G.~F.}\ \bibnamefont {Li}},\
  }\href@noop {} {\  (\bibinfo {year} {2019})},\ \Eprint
  {http://arxiv.org/abs/1908.02312} {arXiv:1908.02312 [gr-qc]} \BibitemShut
  {NoStop}%
\bibitem [{\citenamefont {Fernandez}\ \emph {et~al.}(2019)\citenamefont
  {Fernandez}, \citenamefont {Ghalsasi},\ and\ \citenamefont
  {Profumo}}]{Fernandez2019}%
  \BibitemOpen
  \bibfield  {author} {\bibinfo {author} {\bibfnamefont {N.}~\bibnamefont
  {Fernandez}}, \bibinfo {author} {\bibfnamefont {A.}~\bibnamefont {Ghalsasi}},
  \ and\ \bibinfo {author} {\bibfnamefont {S.}~\bibnamefont {Profumo}},\
  }\href@noop {} {\  (\bibinfo {year} {2019})},\ \Eprint
  {http://arxiv.org/abs/1911.07862} {arXiv:1911.07862 [hep-ph]} \BibitemShut
  {NoStop}%
\bibitem [{\citenamefont {Brenneman}\ \emph {et~al.}(2011)\citenamefont
  {Brenneman}, \citenamefont {Reynolds}, \citenamefont {Nowak}, \citenamefont
  {Reis}, \citenamefont {Trippe} \emph {et~al.}}]{Brenneman:2011wz}%
  \BibitemOpen
  \bibfield  {author} {\bibinfo {author} {\bibfnamefont {L.}~\bibnamefont
  {Brenneman}}, \bibinfo {author} {\bibfnamefont {C.}~\bibnamefont {Reynolds}},
  \bibinfo {author} {\bibfnamefont {M.}~\bibnamefont {Nowak}}, \bibinfo
  {author} {\bibfnamefont {R.}~\bibnamefont {Reis}}, \bibinfo {author}
  {\bibfnamefont {M.}~\bibnamefont {Trippe}},  \emph {et~al.},\ }\href
  {\doibase 10.1088/0004-637X/736/2/103} {\bibfield  {journal} {\bibinfo
  {journal} {Astrophys.J.}\ }\textbf {\bibinfo {volume} {736}},\ \bibinfo
  {pages} {103} (\bibinfo {year} {2011})},\ \Eprint
  {http://arxiv.org/abs/1104.1172} {arXiv:1104.1172 [astro-ph.HE]} \BibitemShut
  {NoStop}%
\bibitem [{\citenamefont {Middleton}(2016)}]{Middleton:2015osa}%
  \BibitemOpen
  \bibfield  {author} {\bibinfo {author} {\bibfnamefont {M.}~\bibnamefont
  {Middleton}},\ }\href {\doibase 10.1007/978-3-662-52859-4_3} {\ ,\ \bibinfo
  {pages} {99} (\bibinfo {year} {2016})},\ \Eprint
  {http://arxiv.org/abs/1507.06153} {arXiv:1507.06153 [astro-ph.HE]}
  \BibitemShut {NoStop}%
\bibitem [{\citenamefont {Abbott}\ \emph {et~al.}(2019)\citenamefont {Abbott}
  \emph {et~al.}}]{LIGOScientific:2018mvr}%
  \BibitemOpen
  \bibfield  {author} {\bibinfo {author} {\bibfnamefont {B.~P.}\ \bibnamefont
  {Abbott}} \emph {et~al.} (\bibinfo {collaboration} {LIGO Scientific,
  Virgo}),\ }\href {\doibase 10.1103/PhysRevX.9.031040} {\bibfield  {journal}
  {\bibinfo  {journal} {Phys. Rev.}\ }\textbf {\bibinfo {volume} {X9}},\
  \bibinfo {pages} {031040} (\bibinfo {year} {2019})},\ \Eprint
  {http://arxiv.org/abs/1811.12907} {arXiv:1811.12907 [astro-ph.HE]}
  \BibitemShut {NoStop}%
\bibitem [{\citenamefont {Venumadhav}\ \emph {et~al.}(2019)\citenamefont
  {Venumadhav}, \citenamefont {Zackay}, \citenamefont {Roulet}, \citenamefont
  {Dai},\ and\ \citenamefont {Zaldarriaga}}]{Venumadhav:2019lyq}%
  \BibitemOpen
  \bibfield  {author} {\bibinfo {author} {\bibfnamefont {T.}~\bibnamefont
  {Venumadhav}}, \bibinfo {author} {\bibfnamefont {B.}~\bibnamefont {Zackay}},
  \bibinfo {author} {\bibfnamefont {J.}~\bibnamefont {Roulet}}, \bibinfo
  {author} {\bibfnamefont {L.}~\bibnamefont {Dai}}, \ and\ \bibinfo {author}
  {\bibfnamefont {M.}~\bibnamefont {Zaldarriaga}},\ }\href@noop {} {\
  (\bibinfo {year} {2019})},\ \Eprint {http://arxiv.org/abs/1904.07214}
  {arXiv:1904.07214 [astro-ph.HE]} \BibitemShut {NoStop}%
\bibitem [{\citenamefont {Abbott}\ \emph {et~al.}(2016)\citenamefont {Abbott}
  \emph {et~al.}}]{TheLIGOScientific:2016pea}%
  \BibitemOpen
  \bibfield  {author} {\bibinfo {author} {\bibfnamefont {B.~P.}\ \bibnamefont
  {Abbott}} \emph {et~al.} (\bibinfo {collaboration} {Virgo, LIGO
  Scientific}),\ }\href {\doibase 10.1103/PhysRevX.6.041015} {\bibfield
  {journal} {\bibinfo  {journal} {Phys. Rev.}\ }\textbf {\bibinfo {volume}
  {X6}},\ \bibinfo {pages} {041015} (\bibinfo {year} {2016})},\ \Eprint
  {http://arxiv.org/abs/1606.04856} {arXiv:1606.04856 [gr-qc]} \BibitemShut
  {NoStop}%
\bibitem [{\citenamefont {Ghosh}\ \emph {et~al.}(2019)\citenamefont {Ghosh},
  \citenamefont {Berti}, \citenamefont {Brito},\ and\ \citenamefont
  {Richartz}}]{Ghosh:2018gaw}%
  \BibitemOpen
  \bibfield  {author} {\bibinfo {author} {\bibfnamefont {S.}~\bibnamefont
  {Ghosh}}, \bibinfo {author} {\bibfnamefont {E.}~\bibnamefont {Berti}},
  \bibinfo {author} {\bibfnamefont {R.}~\bibnamefont {Brito}}, \ and\ \bibinfo
  {author} {\bibfnamefont {M.}~\bibnamefont {Richartz}},\ }\href {\doibase
  10.1103/PhysRevD.99.104030} {\bibfield  {journal} {\bibinfo  {journal} {Phys.
  Rev.}\ }\textbf {\bibinfo {volume} {D99}},\ \bibinfo {pages} {104030}
  (\bibinfo {year} {2019})},\ \Eprint {http://arxiv.org/abs/1812.01620}
  {arXiv:1812.01620 [gr-qc]} \BibitemShut {NoStop}%
\bibitem [{\citenamefont {Cardoso}\ \emph {et~al.}(2018)\citenamefont
  {Cardoso}, \citenamefont {Dias}, \citenamefont {Hartnett}, \citenamefont
  {Middleton}, \citenamefont {Pani},\ and\ \citenamefont
  {Santos}}]{Cardoso:2018tly}%
  \BibitemOpen
  \bibfield  {author} {\bibinfo {author} {\bibfnamefont {V.}~\bibnamefont
  {Cardoso}}, \bibinfo {author} {\bibfnamefont {O.~J.~C.}\ \bibnamefont
  {Dias}}, \bibinfo {author} {\bibfnamefont {G.~S.}\ \bibnamefont {Hartnett}},
  \bibinfo {author} {\bibfnamefont {M.}~\bibnamefont {Middleton}}, \bibinfo
  {author} {\bibfnamefont {P.}~\bibnamefont {Pani}}, \ and\ \bibinfo {author}
  {\bibfnamefont {J.~E.}\ \bibnamefont {Santos}},\ }\href {\doibase
  10.1088/1475-7516/2018/03/043} {\bibfield  {journal} {\bibinfo  {journal}
  {JCAP}\ }\textbf {\bibinfo {volume} {1803}},\ \bibinfo {pages} {043}
  (\bibinfo {year} {2018})},\ \Eprint {http://arxiv.org/abs/1801.01420}
  {arXiv:1801.01420 [gr-qc]} \BibitemShut {NoStop}%
\bibitem [{\citenamefont {Steiner}\ \emph {et~al.}(2010)\citenamefont
  {Steiner}, \citenamefont {McClintock}, \citenamefont {Remillard},
  \citenamefont {Gou}, \citenamefont {Yamada},\ and\ \citenamefont
  {Narayan}}]{Steiner:2010kd}%
  \BibitemOpen
  \bibfield  {author} {\bibinfo {author} {\bibfnamefont {J.~F.}\ \bibnamefont
  {Steiner}}, \bibinfo {author} {\bibfnamefont {J.~E.}\ \bibnamefont
  {McClintock}}, \bibinfo {author} {\bibfnamefont {R.~A.}\ \bibnamefont
  {Remillard}}, \bibinfo {author} {\bibfnamefont {L.}~\bibnamefont {Gou}},
  \bibinfo {author} {\bibfnamefont {S.}~\bibnamefont {Yamada}}, \ and\ \bibinfo
  {author} {\bibfnamefont {R.}~\bibnamefont {Narayan}},\ }\href {\doibase
  10.1088/2041-8205/718/2/L117} {\bibfield  {journal} {\bibinfo  {journal}
  {Astrophys. J.}\ }\textbf {\bibinfo {volume} {718}},\ \bibinfo {pages} {L117}
  (\bibinfo {year} {2010})},\ \Eprint {http://arxiv.org/abs/1006.5729}
  {arXiv:1006.5729 [astro-ph.HE]} \BibitemShut {NoStop}%
\bibitem [{\citenamefont {Gou}\ \emph {et~al.}(2009)\citenamefont {Gou},
  \citenamefont {McClintock}, \citenamefont {Liu}, \citenamefont {Narayan},
  \citenamefont {Steiner}, \citenamefont {Remillard}, \citenamefont {Orosz},\
  and\ \citenamefont {Davis}}]{Gou:2009ks}%
  \BibitemOpen
  \bibfield  {author} {\bibinfo {author} {\bibfnamefont {L.}~\bibnamefont
  {Gou}}, \bibinfo {author} {\bibfnamefont {J.~E.}\ \bibnamefont {McClintock}},
  \bibinfo {author} {\bibfnamefont {J.}~\bibnamefont {Liu}}, \bibinfo {author}
  {\bibfnamefont {R.}~\bibnamefont {Narayan}}, \bibinfo {author} {\bibfnamefont
  {J.~F.}\ \bibnamefont {Steiner}}, \bibinfo {author} {\bibfnamefont {R.~A.}\
  \bibnamefont {Remillard}}, \bibinfo {author} {\bibfnamefont {J.~A.}\
  \bibnamefont {Orosz}}, \ and\ \bibinfo {author} {\bibfnamefont {S.~W.}\
  \bibnamefont {Davis}},\ }\href {\doibase 10.1088/0004-637X/701/2/1076}
  {\bibfield  {journal} {\bibinfo  {journal} {Astrophys. J.}\ }\textbf
  {\bibinfo {volume} {701}},\ \bibinfo {pages} {1076} (\bibinfo {year}
  {2009})},\ \Eprint {http://arxiv.org/abs/0901.0920} {arXiv:0901.0920
  [astro-ph.HE]} \BibitemShut {NoStop}%
\bibitem [{\citenamefont {Akiyama}\ \emph
  {et~al.}(2019{\natexlab{a}})\citenamefont {Akiyama} \emph
  {et~al.}}]{Akiyama:2019cqa}%
  \BibitemOpen
  \bibfield  {author} {\bibinfo {author} {\bibfnamefont {K.}~\bibnamefont
  {Akiyama}} \emph {et~al.} (\bibinfo {collaboration} {Event Horizon
  Telescope}),\ }\href {\doibase 10.3847/2041-8213/ab0ec7} {\bibfield
  {journal} {\bibinfo  {journal} {Astrophys. J.}\ }\textbf {\bibinfo {volume}
  {875}},\ \bibinfo {pages} {L1} (\bibinfo {year}
  {2019}{\natexlab{a}})}\BibitemShut {NoStop}%
\bibitem [{\citenamefont {Akiyama}\ \emph
  {et~al.}(2019{\natexlab{b}})\citenamefont {Akiyama} \emph
  {et~al.}}]{Akiyama:2019eap}%
  \BibitemOpen
  \bibfield  {author} {\bibinfo {author} {\bibfnamefont {K.}~\bibnamefont
  {Akiyama}} \emph {et~al.} (\bibinfo {collaboration} {Event Horizon
  Telescope}),\ }\href {\doibase 10.3847/2041-8213/ab1141} {\bibfield
  {journal} {\bibinfo  {journal} {Astrophys. J.}\ }\textbf {\bibinfo {volume}
  {875}},\ \bibinfo {pages} {L6} (\bibinfo {year} {2019}{\natexlab{b}})},\
  \Eprint {http://arxiv.org/abs/1906.11243} {arXiv:1906.11243 [astro-ph.GA]}
  \BibitemShut {NoStop}%
\bibitem [{\citenamefont {Akiyama}\ \emph
  {et~al.}(2019{\natexlab{c}})\citenamefont {Akiyama} \emph
  {et~al.}}]{Akiyama:2019fyp}%
  \BibitemOpen
  \bibfield  {author} {\bibinfo {author} {\bibfnamefont {K.}~\bibnamefont
  {Akiyama}} \emph {et~al.} (\bibinfo {collaboration} {Event Horizon
  Telescope}),\ }\href {\doibase 10.3847/2041-8213/ab0f43} {\bibfield
  {journal} {\bibinfo  {journal} {Astrophys. J.}\ }\textbf {\bibinfo {volume}
  {875}},\ \bibinfo {pages} {L5} (\bibinfo {year} {2019}{\natexlab{c}})},\
  \Eprint {http://arxiv.org/abs/1906.11242} {arXiv:1906.11242 [astro-ph.GA]}
  \BibitemShut {NoStop}%
\bibitem [{\citenamefont {Tamburini}\ \emph {et~al.}(2019)\citenamefont
  {Tamburini}, \citenamefont {Thid\'{e}},\ and\ \citenamefont
  {Della~Valle}}]{Tamburini:2019vrf}%
  \BibitemOpen
  \bibfield  {author} {\bibinfo {author} {\bibfnamefont {F.}~\bibnamefont
  {Tamburini}}, \bibinfo {author} {\bibfnamefont {B.}~\bibnamefont
  {Thid\'{e}}}, \ and\ \bibinfo {author} {\bibfnamefont {M.}~\bibnamefont
  {Della~Valle}},\ }\href@noop {} {\  (\bibinfo {year} {2019})},\ \Eprint
  {http://arxiv.org/abs/1904.07923} {arXiv:1904.07923 [astro-ph.HE]}
  \BibitemShut {NoStop}%
\bibitem [{\citenamefont {Davoudiasl}\ and\ \citenamefont
  {Denton}(2019)}]{Davoudiasl:2019nlo}%
  \BibitemOpen
  \bibfield  {author} {\bibinfo {author} {\bibfnamefont {H.}~\bibnamefont
  {Davoudiasl}}\ and\ \bibinfo {author} {\bibfnamefont {P.~B.}\ \bibnamefont
  {Denton}},\ }\href {\doibase 10.1103/PhysRevLett.123.021102} {\bibfield
  {journal} {\bibinfo  {journal} {Phys. Rev. Lett.}\ }\textbf {\bibinfo
  {volume} {123}},\ \bibinfo {pages} {021102} (\bibinfo {year} {2019})},\
  \Eprint {http://arxiv.org/abs/1904.09242} {arXiv:1904.09242 [astro-ph.CO]}
  \BibitemShut {NoStop}%
\bibitem [{\citenamefont {Chen}\ \emph {et~al.}(2019)\citenamefont {Chen},
  \citenamefont {Shu}, \citenamefont {Xue}, \citenamefont {Yuan},\ and\
  \citenamefont {Zhao}}]{Chen:2019fsq}%
  \BibitemOpen
  \bibfield  {author} {\bibinfo {author} {\bibfnamefont {Y.}~\bibnamefont
  {Chen}}, \bibinfo {author} {\bibfnamefont {J.}~\bibnamefont {Shu}}, \bibinfo
  {author} {\bibfnamefont {X.}~\bibnamefont {Xue}}, \bibinfo {author}
  {\bibfnamefont {Q.}~\bibnamefont {Yuan}}, \ and\ \bibinfo {author}
  {\bibfnamefont {Y.}~\bibnamefont {Zhao}},\ }\href@noop {} {\  (\bibinfo
  {year} {2019})},\ \Eprint {http://arxiv.org/abs/1905.02213} {arXiv:1905.02213
  [hep-ph]} \BibitemShut {NoStop}%
\bibitem [{\citenamefont {Cunha}\ \emph {et~al.}(2019)\citenamefont {Cunha},
  \citenamefont {Herdeiro},\ and\ \citenamefont {Radu}}]{Cunha:2019ikd}%
  \BibitemOpen
  \bibfield  {author} {\bibinfo {author} {\bibfnamefont {P.~V.~P.}\
  \bibnamefont {Cunha}}, \bibinfo {author} {\bibfnamefont {C.~A.~R.}\
  \bibnamefont {Herdeiro}}, \ and\ \bibinfo {author} {\bibfnamefont
  {E.}~\bibnamefont {Radu}},\ }\href {\doibase 10.3390/universe5120220}
  {\bibfield  {journal} {\bibinfo  {journal} {Universe}\ }\textbf {\bibinfo
  {volume} {5}},\ \bibinfo {pages} {220} (\bibinfo {year} {2019})},\ \Eprint
  {http://arxiv.org/abs/1909.08039} {arXiv:1909.08039 [gr-qc]} \BibitemShut
  {NoStop}%
\bibitem [{\citenamefont {McConnell}\ \emph {et~al.}(2011)\citenamefont
  {McConnell}, \citenamefont {Ma}, \citenamefont {Gebhardt}, \citenamefont
  {Wright}, \citenamefont {Murphy} \emph {et~al.}}]{McConnell:2011mu}%
  \BibitemOpen
  \bibfield  {author} {\bibinfo {author} {\bibfnamefont {N.~J.}\ \bibnamefont
  {McConnell}}, \bibinfo {author} {\bibfnamefont {C.-P.}\ \bibnamefont {Ma}},
  \bibinfo {author} {\bibfnamefont {K.}~\bibnamefont {Gebhardt}}, \bibinfo
  {author} {\bibfnamefont {S.~A.}\ \bibnamefont {Wright}}, \bibinfo {author}
  {\bibfnamefont {J.~D.}\ \bibnamefont {Murphy}},  \emph {et~al.},\ }\href
  {\doibase 10.1038/nature10636} {\bibfield  {journal} {\bibinfo  {journal}
  {Nature}\ }\textbf {\bibinfo {volume} {480}},\ \bibinfo {pages} {215}
  (\bibinfo {year} {2011})},\ \Eprint {http://arxiv.org/abs/1112.1078}
  {arXiv:1112.1078 [astro-ph.CO]} \BibitemShut {NoStop}%
\bibitem [{\citenamefont {McConnell}\ \emph {et~al.}(2012)\citenamefont
  {McConnell}, \citenamefont {Ma}, \citenamefont {Murphy}, \citenamefont
  {Gebhardt}, \citenamefont {Lauer} \emph {et~al.}}]{2012arXiv1203.1620M}%
  \BibitemOpen
  \bibfield  {author} {\bibinfo {author} {\bibfnamefont {N.~J.}\ \bibnamefont
  {McConnell}}, \bibinfo {author} {\bibfnamefont {C.-P.}\ \bibnamefont {Ma}},
  \bibinfo {author} {\bibfnamefont {J.~D.}\ \bibnamefont {Murphy}}, \bibinfo
  {author} {\bibfnamefont {K.}~\bibnamefont {Gebhardt}}, \bibinfo {author}
  {\bibfnamefont {T.~R.}\ \bibnamefont {Lauer}},  \emph {et~al.},\ }\href
  {\doibase 10.1088/0004-637X/756/2/179} {\bibfield  {journal} {\bibinfo
  {journal} {Astrophysical Journal}\ }\textbf {\bibinfo {volume} {756}},\
  \bibinfo {pages} {179} (\bibinfo {year} {2012})},\ \Eprint
  {http://arxiv.org/abs/1203.1620} {arXiv:1203.1620 [astro-ph.CO]} \BibitemShut
  {NoStop}%
\bibitem [{\citenamefont {Riechers}\ \emph {et~al.}(2009)\citenamefont
  {Riechers}, \citenamefont {Walter}, \citenamefont {Carilli},\ and\
  \citenamefont {Lewis}}]{Riechers:2008xt}%
  \BibitemOpen
  \bibfield  {author} {\bibinfo {author} {\bibfnamefont {D.~A.}\ \bibnamefont
  {Riechers}}, \bibinfo {author} {\bibfnamefont {F.}~\bibnamefont {Walter}},
  \bibinfo {author} {\bibfnamefont {C.~L.}\ \bibnamefont {Carilli}}, \ and\
  \bibinfo {author} {\bibfnamefont {G.~F.}\ \bibnamefont {Lewis}},\ }\href
  {\doibase 10.1088/0004-637X/690/1/463} {\bibfield  {journal} {\bibinfo
  {journal} {Astrophys. J.}\ }\textbf {\bibinfo {volume} {690}},\ \bibinfo
  {pages} {463} (\bibinfo {year} {2009})},\ \Eprint
  {http://arxiv.org/abs/0809.0754} {arXiv:0809.0754 [astro-ph]} \BibitemShut
  {NoStop}%
\bibitem [{\citenamefont {Klein}\ \emph {et~al.}(2016)\citenamefont {Klein}
  \emph {et~al.}}]{Klein:2015hvg}%
  \BibitemOpen
  \bibfield  {author} {\bibinfo {author} {\bibfnamefont {A.}~\bibnamefont
  {Klein}} \emph {et~al.},\ }\href {\doibase 10.1103/PhysRevD.93.024003}
  {\bibfield  {journal} {\bibinfo  {journal} {Phys. Rev.}\ }\textbf {\bibinfo
  {volume} {D93}},\ \bibinfo {pages} {024003} (\bibinfo {year} {2016})},\
  \Eprint {http://arxiv.org/abs/1511.05581} {arXiv:1511.05581 [gr-qc]}
  \BibitemShut {NoStop}%
\bibitem [{\citenamefont {Kyutoku}\ \emph {et~al.}(2020)\citenamefont
  {Kyutoku}, \citenamefont {Fujibayashi}, \citenamefont {Hayashi},
  \citenamefont {Kawaguchi}, \citenamefont {Kiuchi}, \citenamefont {Shibata},\
  and\ \citenamefont {Tanaka}}]{Kyutoku:2020xka}%
  \BibitemOpen
  \bibfield  {author} {\bibinfo {author} {\bibfnamefont {K.}~\bibnamefont
  {Kyutoku}}, \bibinfo {author} {\bibfnamefont {S.}~\bibnamefont
  {Fujibayashi}}, \bibinfo {author} {\bibfnamefont {K.}~\bibnamefont
  {Hayashi}}, \bibinfo {author} {\bibfnamefont {K.}~\bibnamefont {Kawaguchi}},
  \bibinfo {author} {\bibfnamefont {K.}~\bibnamefont {Kiuchi}}, \bibinfo
  {author} {\bibfnamefont {M.}~\bibnamefont {Shibata}}, \ and\ \bibinfo
  {author} {\bibfnamefont {M.}~\bibnamefont {Tanaka}},\ }\href@noop {} {\
  (\bibinfo {year} {2020})},\ \Eprint {http://arxiv.org/abs/2001.04474}
  {arXiv:2001.04474 [astro-ph.HE]} \BibitemShut {NoStop}%
\bibitem [{\citenamefont {Berti}\ \emph {et~al.}(2006)\citenamefont {Berti},
  \citenamefont {Cardoso},\ and\ \citenamefont {Will}}]{Berti:2005ys}%
  \BibitemOpen
  \bibfield  {author} {\bibinfo {author} {\bibfnamefont {E.}~\bibnamefont
  {Berti}}, \bibinfo {author} {\bibfnamefont {V.}~\bibnamefont {Cardoso}}, \
  and\ \bibinfo {author} {\bibfnamefont {C.~M.}\ \bibnamefont {Will}},\ }\href
  {\doibase 10.1103/PhysRevD.73.064030} {\bibfield  {journal} {\bibinfo
  {journal} {Phys. Rev.}\ }\textbf {\bibinfo {volume} {D73}},\ \bibinfo {pages}
  {064030} (\bibinfo {year} {2006})},\ \Eprint
  {http://arxiv.org/abs/gr-qc/0512160} {arXiv:gr-qc/0512160 [gr-qc]}
  \BibitemShut {NoStop}%
\bibitem [{\citenamefont {Tsukada}\ \emph {et~al.}(2019)\citenamefont
  {Tsukada}, \citenamefont {Callister}, \citenamefont {Matas},\ and\
  \citenamefont {Meyers}}]{Tsukada:2018mbp}%
  \BibitemOpen
  \bibfield  {author} {\bibinfo {author} {\bibfnamefont {L.}~\bibnamefont
  {Tsukada}}, \bibinfo {author} {\bibfnamefont {T.}~\bibnamefont {Callister}},
  \bibinfo {author} {\bibfnamefont {A.}~\bibnamefont {Matas}}, \ and\ \bibinfo
  {author} {\bibfnamefont {P.}~\bibnamefont {Meyers}},\ }\href {\doibase
  10.1103/PhysRevD.99.103015} {\bibfield  {journal} {\bibinfo  {journal} {Phys.
  Rev.}\ }\textbf {\bibinfo {volume} {D99}},\ \bibinfo {pages} {103015}
  (\bibinfo {year} {2019})},\ \Eprint {http://arxiv.org/abs/1812.09622}
  {arXiv:1812.09622 [astro-ph.HE]} \BibitemShut {NoStop}%
\bibitem [{\citenamefont {Sun}\ \emph {et~al.}(2019)\citenamefont {Sun},
  \citenamefont {Brito},\ and\ \citenamefont {Isi}}]{Sun:2019mqb}%
  \BibitemOpen
  \bibfield  {author} {\bibinfo {author} {\bibfnamefont {L.}~\bibnamefont
  {Sun}}, \bibinfo {author} {\bibfnamefont {R.}~\bibnamefont {Brito}}, \ and\
  \bibinfo {author} {\bibfnamefont {M.}~\bibnamefont {Isi}},\ }\href@noop {} {\
   (\bibinfo {year} {2019})},\ \Eprint {http://arxiv.org/abs/1909.11267}
  {arXiv:1909.11267 [gr-qc]} \BibitemShut {NoStop}%
\bibitem [{\citenamefont {Zhu}\ \emph {et~al.}(2020)\citenamefont {Zhu},
  \citenamefont {Baryakhtar}, \citenamefont {Papa}, \citenamefont {Tsuna},
  \citenamefont {Kawanaka},\ and\ \citenamefont {Eggenstein}}]{Zhu:2020tht}%
  \BibitemOpen
  \bibfield  {author} {\bibinfo {author} {\bibfnamefont {S.~J.}\ \bibnamefont
  {Zhu}}, \bibinfo {author} {\bibfnamefont {M.}~\bibnamefont {Baryakhtar}},
  \bibinfo {author} {\bibfnamefont {M.~A.}\ \bibnamefont {Papa}}, \bibinfo
  {author} {\bibfnamefont {D.}~\bibnamefont {Tsuna}}, \bibinfo {author}
  {\bibfnamefont {N.}~\bibnamefont {Kawanaka}}, \ and\ \bibinfo {author}
  {\bibfnamefont {H.-B.}\ \bibnamefont {Eggenstein}},\ }\href@noop {} {\
  (\bibinfo {year} {2020})},\ \Eprint {http://arxiv.org/abs/2003.03359}
  {arXiv:2003.03359 [gr-qc]} \BibitemShut {NoStop}%
\bibitem [{\citenamefont {Hannuksela}\ \emph {et~al.}(2019)\citenamefont
  {Hannuksela}, \citenamefont {Wong}, \citenamefont {Brito}, \citenamefont
  {Berti},\ and\ \citenamefont {Li}}]{Hannuksela:2018izj}%
  \BibitemOpen
  \bibfield  {author} {\bibinfo {author} {\bibfnamefont {O.~A.}\ \bibnamefont
  {Hannuksela}}, \bibinfo {author} {\bibfnamefont {K.~W.~K.}\ \bibnamefont
  {Wong}}, \bibinfo {author} {\bibfnamefont {R.}~\bibnamefont {Brito}},
  \bibinfo {author} {\bibfnamefont {E.}~\bibnamefont {Berti}}, \ and\ \bibinfo
  {author} {\bibfnamefont {T.~G.~F.}\ \bibnamefont {Li}},\ }\href {\doibase
  10.1038/s41550-019-0712-4} {\bibfield  {journal} {\bibinfo  {journal} {Nat.
  Astron.}\ }\textbf {\bibinfo {volume} {3}},\ \bibinfo {pages} {447} (\bibinfo
  {year} {2019})},\ \Eprint {http://arxiv.org/abs/1804.09659} {arXiv:1804.09659
  [astro-ph.HE]} \BibitemShut {NoStop}%
\bibitem [{\citenamefont {Baumann}\ \emph
  {et~al.}(2019{\natexlab{a}})\citenamefont {Baumann}, \citenamefont {Chia},\
  and\ \citenamefont {Porto}}]{Baumann:2018vus}%
  \BibitemOpen
  \bibfield  {author} {\bibinfo {author} {\bibfnamefont {D.}~\bibnamefont
  {Baumann}}, \bibinfo {author} {\bibfnamefont {H.~S.}\ \bibnamefont {Chia}}, \
  and\ \bibinfo {author} {\bibfnamefont {R.~A.}\ \bibnamefont {Porto}},\ }\href
  {\doibase 10.1103/PhysRevD.99.044001} {\bibfield  {journal} {\bibinfo
  {journal} {Phys. Rev.}\ }\textbf {\bibinfo {volume} {D99}},\ \bibinfo {pages}
  {044001} (\bibinfo {year} {2019}{\natexlab{a}})},\ \Eprint
  {http://arxiv.org/abs/1804.03208} {arXiv:1804.03208 [gr-qc]} \BibitemShut
  {NoStop}%
\bibitem [{\citenamefont {Zhang}\ and\ \citenamefont
  {Yang}(2019{\natexlab{a}})}]{Zhang:2018kib}%
  \BibitemOpen
  \bibfield  {author} {\bibinfo {author} {\bibfnamefont {J.}~\bibnamefont
  {Zhang}}\ and\ \bibinfo {author} {\bibfnamefont {H.}~\bibnamefont {Yang}},\
  }\href {\doibase 10.1103/PhysRevD.99.064018} {\bibfield  {journal} {\bibinfo
  {journal} {Phys. Rev.}\ }\textbf {\bibinfo {volume} {D99}},\ \bibinfo {pages}
  {064018} (\bibinfo {year} {2019}{\natexlab{a}})},\ \Eprint
  {http://arxiv.org/abs/1808.02905} {arXiv:1808.02905 [gr-qc]} \BibitemShut
  {NoStop}%
\bibitem [{\citenamefont {Berti}\ \emph {et~al.}(2019)\citenamefont {Berti},
  \citenamefont {Brito}, \citenamefont {Macedo}, \citenamefont {Raposo},\ and\
  \citenamefont {Rosa}}]{Berti:2019wnn}%
  \BibitemOpen
  \bibfield  {author} {\bibinfo {author} {\bibfnamefont {E.}~\bibnamefont
  {Berti}}, \bibinfo {author} {\bibfnamefont {R.}~\bibnamefont {Brito}},
  \bibinfo {author} {\bibfnamefont {C.~F.~B.}\ \bibnamefont {Macedo}}, \bibinfo
  {author} {\bibfnamefont {G.}~\bibnamefont {Raposo}}, \ and\ \bibinfo {author}
  {\bibfnamefont {J.~L.}\ \bibnamefont {Rosa}},\ }\href {\doibase
  10.1103/PhysRevD.99.104039} {\bibfield  {journal} {\bibinfo  {journal} {Phys.
  Rev.}\ }\textbf {\bibinfo {volume} {D99}},\ \bibinfo {pages} {104039}
  (\bibinfo {year} {2019})},\ \Eprint {http://arxiv.org/abs/1904.03131}
  {arXiv:1904.03131 [gr-qc]} \BibitemShut {NoStop}%
\bibitem [{\citenamefont {Baumann}\ \emph
  {et~al.}(2019{\natexlab{b}})\citenamefont {Baumann}, \citenamefont {Chia},
  \citenamefont {Stout},\ and\ \citenamefont {ter Haar}}]{Baumann:2019eav}%
  \BibitemOpen
  \bibfield  {author} {\bibinfo {author} {\bibfnamefont {D.}~\bibnamefont
  {Baumann}}, \bibinfo {author} {\bibfnamefont {H.~S.}\ \bibnamefont {Chia}},
  \bibinfo {author} {\bibfnamefont {J.}~\bibnamefont {Stout}}, \ and\ \bibinfo
  {author} {\bibfnamefont {L.}~\bibnamefont {ter Haar}},\ }\href@noop {} {\
  (\bibinfo {year} {2019}{\natexlab{b}})},\ \Eprint
  {http://arxiv.org/abs/1908.10370} {arXiv:1908.10370 [gr-qc]} \BibitemShut
  {NoStop}%
\bibitem [{\citenamefont {Baumann}\ \emph
  {et~al.}(2019{\natexlab{c}})\citenamefont {Baumann}, \citenamefont {Chia},
  \citenamefont {Porto},\ and\ \citenamefont {Stout}}]{Baumann:2019ztm}%
  \BibitemOpen
  \bibfield  {author} {\bibinfo {author} {\bibfnamefont {D.}~\bibnamefont
  {Baumann}}, \bibinfo {author} {\bibfnamefont {H.~S.}\ \bibnamefont {Chia}},
  \bibinfo {author} {\bibfnamefont {R.~A.}\ \bibnamefont {Porto}}, \ and\
  \bibinfo {author} {\bibfnamefont {J.}~\bibnamefont {Stout}},\ }\href@noop {}
  {\  (\bibinfo {year} {2019}{\natexlab{c}})},\ \Eprint
  {http://arxiv.org/abs/1912.04932} {arXiv:1912.04932 [gr-qc]} \BibitemShut
  {NoStop}%
\bibitem [{\citenamefont {Zhang}\ and\ \citenamefont
  {Yang}(2019{\natexlab{b}})}]{Zhang:2019eid}%
  \BibitemOpen
  \bibfield  {author} {\bibinfo {author} {\bibfnamefont {J.}~\bibnamefont
  {Zhang}}\ and\ \bibinfo {author} {\bibfnamefont {H.}~\bibnamefont {Yang}},\
  }\href@noop {} {\  (\bibinfo {year} {2019}{\natexlab{b}})},\ \Eprint
  {http://arxiv.org/abs/1907.13582} {arXiv:1907.13582 [gr-qc]} \BibitemShut
  {NoStop}%
\bibitem [{\citenamefont {Cardoso}\ \emph {et~al.}(2020)\citenamefont
  {Cardoso}, \citenamefont {Duque},\ and\ \citenamefont
  {Ikeda}}]{Cardoso:2020hca}%
  \BibitemOpen
  \bibfield  {author} {\bibinfo {author} {\bibfnamefont {V.}~\bibnamefont
  {Cardoso}}, \bibinfo {author} {\bibfnamefont {F.}~\bibnamefont {Duque}}, \
  and\ \bibinfo {author} {\bibfnamefont {T.}~\bibnamefont {Ikeda}},\
  }\href@noop {} {\  (\bibinfo {year} {2020})},\ \Eprint
  {http://arxiv.org/abs/2001.01729} {arXiv:2001.01729 [gr-qc]} \BibitemShut
  {NoStop}%
\bibitem [{\citenamefont {Ikeda}\ \emph {et~al.}(2019)\citenamefont {Ikeda},
  \citenamefont {Brito},\ and\ \citenamefont {Cardoso}}]{Ikeda:2019fvj}%
  \BibitemOpen
  \bibfield  {author} {\bibinfo {author} {\bibfnamefont {T.}~\bibnamefont
  {Ikeda}}, \bibinfo {author} {\bibfnamefont {R.}~\bibnamefont {Brito}}, \ and\
  \bibinfo {author} {\bibfnamefont {V.}~\bibnamefont {Cardoso}},\ }\href
  {\doibase 10.1103/PhysRevLett.122.081101} {\bibfield  {journal} {\bibinfo
  {journal} {Phys. Rev. Lett.}\ }\textbf {\bibinfo {volume} {122}},\ \bibinfo
  {pages} {081101} (\bibinfo {year} {2019})},\ \Eprint
  {http://arxiv.org/abs/1811.04950} {arXiv:1811.04950 [gr-qc]} \BibitemShut
  {NoStop}%
\bibitem [{\citenamefont {Yoshino}\ and\ \citenamefont
  {Kodama}(2012)}]{Yoshino:2012kn}%
  \BibitemOpen
  \bibfield  {author} {\bibinfo {author} {\bibfnamefont {H.}~\bibnamefont
  {Yoshino}}\ and\ \bibinfo {author} {\bibfnamefont {H.}~\bibnamefont
  {Kodama}},\ }\href {\doibase 10.1143/PTP.128.153} {\bibfield  {journal}
  {\bibinfo  {journal} {Prog.Theor.Phys.}\ }\textbf {\bibinfo {volume} {128}},\
  \bibinfo {pages} {153} (\bibinfo {year} {2012})},\ \Eprint
  {http://arxiv.org/abs/1203.5070} {arXiv:1203.5070 [gr-qc]} \BibitemShut
  {NoStop}%
\bibitem [{\citenamefont {Yoshino}\ and\ \citenamefont
  {Kodama}(2015)}]{Yoshino:2015nsa}%
  \BibitemOpen
  \bibfield  {author} {\bibinfo {author} {\bibfnamefont {H.}~\bibnamefont
  {Yoshino}}\ and\ \bibinfo {author} {\bibfnamefont {H.}~\bibnamefont
  {Kodama}},\ }\href {\doibase 10.1088/0264-9381/32/21/214001} {\bibfield
  {journal} {\bibinfo  {journal} {Class. Quant. Grav.}\ }\textbf {\bibinfo
  {volume} {32}},\ \bibinfo {pages} {214001} (\bibinfo {year} {2015})},\
  \Eprint {http://arxiv.org/abs/1505.00714} {arXiv:1505.00714 [gr-qc]}
  \BibitemShut {NoStop}%
\end{thebibliography}%

\newpage
\appendix

\section{Massive spin-2 instability and GW emission in the small-coupling limit} \label{app:SM}
In this appendix we discuss in some detail the procedure to solve for the massive spin-$2$ field equations in a BH 
background and compute the GW signal from the massive spin-$2$ condensate in the small-coupling ($\alpha=M\mu\ll1$) 
limit. We assume a Kerr background metric but remark that the details of the background are not relevant in the 
small-coupling limit, and the procedure should also work for different backgrounds that might exist in bimetric 
theories~\cite{Babichev:2015xha}.
It is remarkable that an analytical solution exists in this limit, despite the fact that it is unknown 
whether the field equations of a massive spin-$2$ field are separable on a Kerr metric for generic couplings. 

\subsection{Hydrogenic solutions in the far-zone}\label{sec:hydrogen_sol}
We first consider the regime where $M\mu\ll 1$ and 
$r\gg M$, but where $r$ is not necessarily large when compared to $1/\mu$. In addition, to be able to 
solve the field equations analytically we also need to consider the region where the Riemann tensor term is much 
smaller than the mass term. To leading order in an expansion in $M/r$, $R_{abcd}\sim\mathcal{O}(M/r^3)$. Therefore, to 
neglect the coupling to the Riemann tensor we require 
$r\gg 
(M/\mu^2)^{1/3}$, which can be rewritten as $(r\mu)^2\gg M/r$. Since $M/r \ll 1$ such region exists as long as $r\mu$ 
is not too small.

We now show that within these approximations, the field equations can be written as a set of non-relativistic 
Schr\"{o}dinger equations with a $1/r$ potential. The derivation follows closely Ref.~\cite{Aoki:2017ixz} and 
generalizes the calculations of Ref.~\citep{Baryakhtar:2017ngi}, where a similar derivation for massive vector fields 
was presented. In the non-relativistic limit the background  metric can be written as 
\begin{equation}
\dif s^{2} = -(1+2\Phi)dt^2 + (1+2\Psi)\gamma_{i j}dx^i\,dx^j\,,
\end{equation}
where $\Phi=-\Psi=-M/r$ and $\gamma_{i j}$ is the 3-dimensional Euclidean metric. Indices $i,j$ are raised 
and lowered by $\gamma_{i j}$. The spin-2 field can be generically written as 
\begin{equation}\label{eq:hmunu}
H_{ab}=\frac{1}{\sqrt{2\mu}}e^{-i\omega t}\begin{pmatrix}
H_{00} & H_{0i} \\
* & \frac{\psi_{\rm tr}}{3}\gamma_{i j}+\psi_{ij}
\end{pmatrix}\,,
\end{equation}
where $H_{00},H_{0i}, \psi_{\rm tr}$, and $\psi_{ij}$ are functions of $(t,x^i)$, and $\psi_{ij}$ is traceless. We 
assume 
that these functions vary on time scales much longer than $\omega^{-1}$, such that, e.g, $\partial^2_{t}\psi_{ij}\ll 
\omega\partial_{t}\psi_{ij}$ or $\partial_{k}\partial^{k}\psi_{ij}\ll \omega\partial_{k}\psi_{ij}$ . Under this 
approximation we obtain from the constraints equations: 
\begin{equation}\label{eq:hmunu_cons}
\psi_{\rm tr}\simeq H_{00}\,,\quad H_{00}\simeq\frac{i}{\omega}\partial^i H_{0i}\,, \quad
H_{0i}\simeq\frac{i}{\omega}\partial^j 
\psi_{ij}\,,
\end{equation}
and therefore one can find $\psi_{\rm tr}, H_{00}$ and $  H_{0i}$ by solving the field equations for $\psi_{ij}$. 
In the non-relativistic approximation this implies $|H_{00}|,|\psi_{\rm tr}| \ll |H_{0i}|\ll |\psi_{ij}|$. 
The equations of motion can be approximated as
\begin{equation}
\left(1+\frac{2M}{r}\right)\omega^2 \psi_{ij}+\nabla^2 \psi_{ij} -\mu^2 \psi_{ij}\simeq0\,.
\end{equation}
For bound states we expect $\omega\sim \mu$. By expanding $\omega$ around $\mu$ and taking the limit $r\gg M$, we find that $\psi_{ij}$ satisfies a set of Schr\"{o}dinger-like equations for a $1/r$ potential:
\begin{equation}\label{eq:schro}
(\omega-\mu) \psi_{ij}\simeq -\frac{\nabla^2 \psi_{ij}}{2\mu}-\frac{\alpha}{r}\psi_{ij}\,,
\end{equation}
%
Thus, the coupling $\alpha$ plays the same role as the fine-structure 
constant in the hydrogen atom. Since the potential is spherically symmetric, we can separate $\psi_{ij}$ into radial 
and angular functions:
\begin{equation}\label{eq:ansatz}
\psi_{ik}=R^{n\ell}(r)Y^{\ell,jm}_{ik}(\theta,\phi)\,,
\end{equation}
where $Y^{\ell,jm}_{ik}(\theta,\phi)$ are eigenfunctions of the orbital 
angular momentum operator
\begin{equation}\label{eq:angular}
-r^2\nabla^2Y^{\ell,jm}_{ik}=\ell(\ell+1)Y^{\ell,jm}_{ik}\,,
\end{equation}
and are therefore called ``pure-orbital'' tensor spherical harmonics. These are defined by (see, 
e.g., Ref.~\cite{Thorne:1980ru} and Chapter~3 of~\cite{Maggiore:1900zz})
\begin{equation}\label{eq:solang}
Y^{\ell,jm}_{ik}(\theta,\phi)=\sum_{\ell_z=-\ell}^{\ell}\sum_{s_z=-2}^{2}\left\langle 2 \ell s_z \ell_z | j m 
\right\rangle Y_{\ell\ell_z}(\theta,\phi)t^{(s_z)}_{ik}\,.
\end{equation}
Here $Y_{\ell\ell_z}(\theta,\phi)$ are the usual scalar spherical harmonics and $t^{(s_z)}_{ik}$ is a traceless 
symmetric tensor defined by 
\begin{equation}
t^{(s_z)}_{ik}=\sum_{m_1,m_2=-1}^{1}\left\langle 1 1 m_1 m_2 | 2 s_z \right\rangle \xi_i^{(m_1)} \xi_k^{(m_2)}\,,
\end{equation}
where the vector $\xi_i^{(s_z)}$ is defined for $s_z=0,\pm 1$ and can be constructed from the unit Cartesian vectors 
$\bm{e}_{x}$, $\bm{e}_{y}$ and $\bm{e}_{z}$:
\begin{equation}
\bm{\xi}^{(\pm 1)}=\mp \frac{1}{\sqrt{2}}\left(\bm{e}_{x}\pm i\bm{e}_{y} \right)\,,\qquad \bm{\xi}^{(0)}=\bm{e}_{z}\,.
\end{equation}

On the other hand, using Eq.~\eqref{eq:angular} one finds that the radial function $R^{n\ell}(r)$ satisfies a 
hydrogen-like radial equation 
\begin{equation}\label{eq:radial}
-\frac{1}{2\mu r^2}\frac{d}{dr}\left(r^2 \frac{d}{dr} R^{n\ell}\right)+\frac{\ell(\ell+1)}{2\mu 
r^2}R^{n\ell}-\frac{\alpha}{r}R^{n\ell}=(\omega-\mu)R^{n\ell}\,.
\end{equation}
Thus, the radial wave functions are hydrogenic, 
labeled by their orbital angular moment $\ell$ and by the overtone number $n$, the latter representing the number of 
nodes in the radial function. Requiring regularity at $r\to \infty$ and at $r=0$ we find that the energy levels are 
given 
by\footnote{Note 
that in most textbooks the hydrogen atom energy levels are written in terms of the principal quantum number 
$\bar{n}\equiv \ell+n+1$ and the energy levels are represented in terms of the bound-state's binding energy 
$E_{\bar{n}}\equiv 
\omega-\mu=-\frac{\alpha^2\mu}{2\bar{n}^2}$. Here we will instead still write energy levels in terms of $\omega$ and 
overtone number $n$ since this is standard notation used in the superradiance literature.} 
\begin{equation}
\omega_R\simeq\mu\left(1-\frac{\alpha^2}{2\left(\ell+n+1\right)^2}\right)\,, \label{wRapp}
\end{equation}
and that the regular solution to Eq.~\eqref{eq:radial} can be written in terms of generalized Laguerre polynomials:
\begin{equation}\label{eq:solradial}
R^{n\ell}=N_{n\ell}\tilde{r}^{\ell}e^{-\tilde{r}/2}L_{n}^{(2\ell+1)}\left(\tilde{r}\right)\,,
\end{equation}
where we defined $\tilde{r}\equiv 2r\alpha\mu/\left(\ell+n+1\right)$ and $N_{n\ell}$ is an arbitrary normalization 
constant which we fix by requiring $\int_0^{\infty} r^2 |R^{n\ell}|^2dr =1$.

In the $r \gg M$ limit the bound-state solutions of a massive spin-2 field in a BH spacetime can be obtained 
by plugging the radial function~\eqref{eq:solradial} and angular functions~\eqref{eq:solang} into 
Eqs.~\eqref{eq:ansatz} 
and~\eqref{eq:hmunu_cons}, which can then be used to reconstruct the field $H_{ab}$ using Eq.~\eqref{eq:hmunu}.
Note that the eigenfunctions peak~\cite{Brito:2015oca} at $r\sim M/\alpha^2$, the largest length scale in the 
problem.

\subsection{Massless solutions in the near-zone}\label{sec:massless_sol}

In the regime where $r \ll \mu^{-1}$, and assuming eigenfunctions with negligible support when $r\lesssim r_C$, the 
mass term and Riemann tensor term in the field equations are subleading. In this regime the field 
equations are close to the ones of a massless spin-2 field in flat space, supplemented by the constraint 
equations. 

To separate the solutions with different spin projections, it is convenient to write them in terms of ``pure-spin'' 
tensor spherical harmonics. The latter can be written as a linear combination of the pure-orbital tensor harmonics and 
have the convenient property of being appropriate to describe pure-spin states of radially propagating GWs in the most 
general metric theory of gravity (see e.g.~\cite{Thorne:1980ru} and Chapter~3 of~\cite{Maggiore:1900zz}). The pure-spin 
tensor spherical harmonics are 
given by\footnote{For ease of notation we drop the space indices $(i,j)$ and recall the reader that all these quantities 
are 
purely-spatial tensors.} 
\begin{eqnarray}
\bm{Y}^{S0}_{jm}&=& a_{11}\bm{Y}^{j+2,jm}+a_{12}\bm{Y}^{j,jm}+a_{13}\bm{Y}^{j-2,jm} \,,\\
\label{eq:YE1}
\bm{Y}^{E1}_{jm}&=& a_{21}\bm{Y}^{j+2,jm}+a_{22}\bm{Y}^{j,jm}+a_{23}\bm{Y}^{j-2,jm} \,,\\
\label{eq:YE2}
\bm{Y}^{E2}_{jm}&=& a_{31}\bm{Y}^{j+2,jm}+a_{32}\bm{Y}^{j,jm}+a_{33}\bm{Y}^{j-2,jm} \,,\\
\label{eq:YB1}
\bm{Y}^{B1}_{jm}&=& b_{11}i\bm{Y}^{j+1,jm}+b_{12}i\bm{Y}^{j-1,jm} \,,\\
\bm{Y}^{B2}_{jm}&=& b_{21}i\bm{Y}^{j+1,jm}+b_{22}i\bm{Y}^{j-1,jm} \,,
\end{eqnarray}
where $a_{kp}$ and $b_{kp}$ are coefficients that only depend on $j$; for explicit expressions see Table~3.1 in 
Ref.~\cite{Maggiore:1900zz} or Ref.~\cite{Thorne:1980ru}. We note that by construction  
$\bm{Y}^{E1}_{lm}$ and $\bm{Y}^{B1}_{lm}$ vanish for $j=0$, whereas $\bm{Y}^{E2}_{lm}$ and $\bm{Y}^{B2}_{lm}$ vanish 
for 
$j=0$ and $j=1$. Furthemore, all these tensors are symmetric and traceless, while
$\bm{Y}^{E2}_{lm}$ and $\bm{Y}^{B2}_{lm}$ are also transverse. 
For a massless spin-2 field the latter two components are 
appropriate to describe the two dynamical degrees of freedom, since all the other components can be eliminated by 
an appropriate gauge choice. On the other hand, for a massive spin-2 field all five tensor 
harmonics defined above are required to describe the five dynamical degrees of freedom. 

Using the pure-spin tensor spherical harmonics, generic solutions to the massless wave equation can be found by 
multiplying each coefficient $a_{kp}$, $b_{kp}$ by some radial function (see e.g. Ref.~\cite{Mathews:1962}) to be found 
by imposing the massless spin-2 field equations in flat spacetime. 

\subsection{Matching}
In the region $r_C \ll r \ll 1/\mu$ we can match the far-zone solution with the near-zone solution.
Since the matching is done in the spherically-symmetric case, we can set $m=0$ without loss of generality.

Since $|H_{00}|,|\psi_{tr}| \ll |H_{0i}| \ll |\psi_{ik}|$, we neglect the sub-leading components. Therefore, for 
the matching and the decay rate calculation we only consider the contribution from $\psi_{ik}$, and check a posteriori the validity of this assumption by comparing the analytical results with the exact numerical calculations~\cite{Brito:2013wya}. Therefore, in the far zone the 
hydrogenic solution is given by 
\begin{equation}
\label{equ:psifar}
    H_{ik}^{\rm far} = \frac{C_{\rm far}}{\sqrt{2\mu}} \left[ 
N_{n\ell}\tilde{r}^{\ell}e^{-\tilde{r}/2}L_{n}^{(2\ell+1)}\left(\tilde{r}\right) Y_{ik}^{\ell,jm}  e^{-i\omega t} + 
c.c.\right]\;,
\end{equation} 
where $C_{\rm far}$ is a free constant.
While our matching procedure is general, below we shall specialize the calculation for the most relevant 
unstable modes, in particular the quadrupole $j=2$, $\ell=0$ and the dipole $j=\ell=1$.

\subsubsection{Quadrupole case: $j=2$, $\ell = 0$ }
Using the pure-spin tensor spherical harmonic in Eq.~\eqref{eq:YE2} and the procedure outlined above to obtain the 
solution of the massless wave equation, for $j=2$, $\ell=0$ we get
\begin{eqnarray}
    H_{ik}^{\rm near} &=& C_{\rm near} \left[ \left(j_4(\omega r)  Y_{ik}^{4,20} - 2 \sqrt{5}
    j_2(\omega r)  Y_{ik}^{2,20} \right. \right.\nonumber\\
    &&\left.\left.+ \sqrt{14}
    j_0(\omega r)  Y_{ik}^{0,20}\right)e^{-i\omega t} + c.c.\right] \;,
\end{eqnarray}
where $C_{\rm near}$ is a free constant and $j_a$ are the spherical Bessel functions of the first kind of order $a$. 
By using the Taylor expansion of the latter near the origin,
\begin{equation}
    j_a(\omega r) \simeq \frac{1}{(2a+1)!!} (\omega r)^a \;,
\end{equation}
we can match the expanded near-zone solution with a series expansion of 
Eq.~\eqref{equ:psifar} at $r\sim0$.
This matching yields $C_{\rm near} = \frac{1}{\sqrt{7}}\; \alpha^{3/2}\mu\; C_{\rm far}$.

\subsubsection{Dipole case: $j=\ell = 1$}
Similarly, using the pure-spin tensor harmonic in Eq.~\eqref{eq:YE1} and following again the same procedure to find a solution, for $j=\ell=1$, we find
\begin{eqnarray}
    H_{ik}^{\rm near} &=& C_{\rm near} \left[ \left(\sqrt{{2}}j_3(\omega r)  Y_{ik}^{3,10}\right.\right.\nonumber\\
    &&\left.\left.- 
\sqrt{3} 
j_1(\omega r)  Y_{ik}^{1,10} \right)e^{-i\omega t} + c.c.\right] \;.
\end{eqnarray}
In this case the matching procedure yields $C_{\rm near}$ = $-\frac{1}{4}  
\alpha^{5/2}\mu C_{\rm far}$. We note that the full massless solution for $j=\ell=1$ also has components $H^{\rm 
near}_{0i}$ but as argued above those components give sub-leading contributions to the decay rate, and we can therefore 
neglect them.

\subsubsection{Magnetic dipole case: $j=1$, $\ell=2$}

Using a different method than the one employed in this paper, an analytical formula for the decay rate of the $j=1$, 
$\ell=2$ mode in the small $\alpha$ limit and for nonspinning BHs was obtained in Ref.~\cite{Brito:2013wya}. Therefore, 
it is instructive to also consider this (subleading) mode to check the correctness of our computation.

Since this is a magnetic (i.e., odd parity) mode, we can use the pure-spin tensor harmonic in Eq.~\eqref{eq:YB1} to 
find the solution:
\begin{eqnarray}
    H_{ik}^{\rm near} &=& C_{\rm near} \left[ j_2(\omega r) i Y_{ik}^{2,10}e^{-i\omega t}+ c.c.\right] \;.
\end{eqnarray}
In this case the matching procedure yields $C_{\rm near}$ = $-2\sqrt{5} 
\alpha^{7/2}\mu C_{\rm far}/(27\sqrt{3})$.

\subsection{Computation of the instability time scale}
Having found the complete solution we can now compute the instability time scale $\tau_{\rm inst}$.
We shall follow Ref.~\cite{Baryakhtar:2017ngi} and compute the decay rate $\Gamma=1/\tau_{\rm inst}$ of a given mode.

At large radii we can approximate the near-zone solutions as a superposition of ingoing and outgoing massless waves 
using:
\begin{equation}
 j_a(\omega r) \simeq \frac{e^{i\omega r-(a+1)i\pi/2}+e^{-i\omega r+(a+1)i\pi/2}}{2\omega r} \;.
\end{equation}
Assuming that the ingoing waves travel down to the horizon without being back scattered\footnote{This requires that the 
potential felt by the field vanishes close to the BH horizon. This seems to be the case for the modes we consider 
since in the vicinity of the BH horizon the solutions can be written as massless plane 
waves~\cite{Brito:2013wya}.} we can approximate the energy flux through the BH horizon as being given by the energy flux of the ingoing wave multiplied by the BH absorption probability of a long wavelength massless wave with spin-$s$ and total angular momentum $j$
 \begin{eqnarray}
    \mathbb{P}_{abs} &\simeq& 2 \left( \frac{(j-s)!(j+s)!}{(2j)!(2j+1)!!}\right)^2 \left( \frac{\omega_{R} - 
m\Omega_{\rm H}}{{\hat\kappa}}\right)\nonumber\\
&\times&\left( \frac{A_H {\hat\kappa}}{2\pi} \omega \right)^{2j+1} \prod_{q=1}^{j} \left[ 1 + 
\left( 
\frac{\omega_{R} - m\Omega_{\rm H}}{q{\hat\kappa}}\right)^2\right] \,,\nonumber\\ \label{Pabs}
 \end{eqnarray}
with $A_{H}=8 \pi M r_{+}$, $\Omega_{\rm H}=\frac{a}{2Mr_{+}}$, and ${\hat\kappa}=\frac{4\pi(r_{+}-M)}{A_{H}}$.

Using ingoing Eddington-Finkelstein coordinates and the ingoing wave solution, we can compute the energy flux across 
the BH horizon through the stress-energy tensor as
\begin{equation}
 \dot E_H  = \int  T_{ab} k^{a}\xi^{b}\, dA_{H} \;,
\end{equation}
where $k^a$ is the time-like Killing vector of the Schwarzschild metric and $\xi^{a}$ is the Killing vector normal to 
the BH horizon (for Schwarzschild $\xi^{a}=k^a$). This provides the rate of change of the total energy of the cloud 
\begin{equation}\label{energy_condensate}
 M_c=-\int d^3 x \sqrt{-g} T^{0}_{0}\,,
\end{equation}
where $T^{0}_{0}$ is the density of the hydrogenic solution~\eqref{equ:psifar}. The massive spin-2 field stress-energy tensor can be derived from the action given in the main text and is given by~\cite{Babichev:2016bxi,Aoki:2017ixz}
\begin{eqnarray}
 T_{ab}&=& -\frac{1}{16\pi}\left[\frac{1}{2}\nabla_a H^{cd}\nabla_b H_{cd}+\nabla^c H_a^d \left(\nabla_c H_{bd}- \nabla_d H_{bc} \right)
  \right.\nonumber\\
&+& \left. H^{cd}\left(\nabla_a \nabla_b H_{cd} + \nabla_d \nabla_c H_{ab} -2\nabla_d \nabla_{(a} H_{b)c} 
\right) \right.\nonumber\\
&+& \left. \frac{g_{ab}}{2}\left(\nabla_e H^{fc}\nabla_f H^e_c-\frac{3}{2}\nabla_e H_{cd}\nabla^e 
H^{cd}\right)\right.\nonumber\\
&+& \left. \mu^2 \left(H_{da}H_b^d-\frac{3}{4}g_{ab} H^{cd}H_{cd}\right)\right]\,. \label{Tmunu}
\end{eqnarray}

To leading order in $\alpha$, for the dominant unstable mode with $n=0$, 
$\ell=0$ and $j=2$, we obtain, after setting $s=2$ in~\eqref{Pabs}:
\begin{equation}
 \Gamma_{|{\chi}=0}^{\ell=0,j=2} = \frac{\langle\dot{E}_H\rangle}{\langle M_c \rangle} \mathbb{P}_{|{\chi}=0} \simeq 
-\frac{128}{45} \alpha^9 \mu \;,
\end{equation}
where the angle brackets indicate a time average. 

\begin{figure*}[t]
\includegraphics[width=0.49\textwidth]{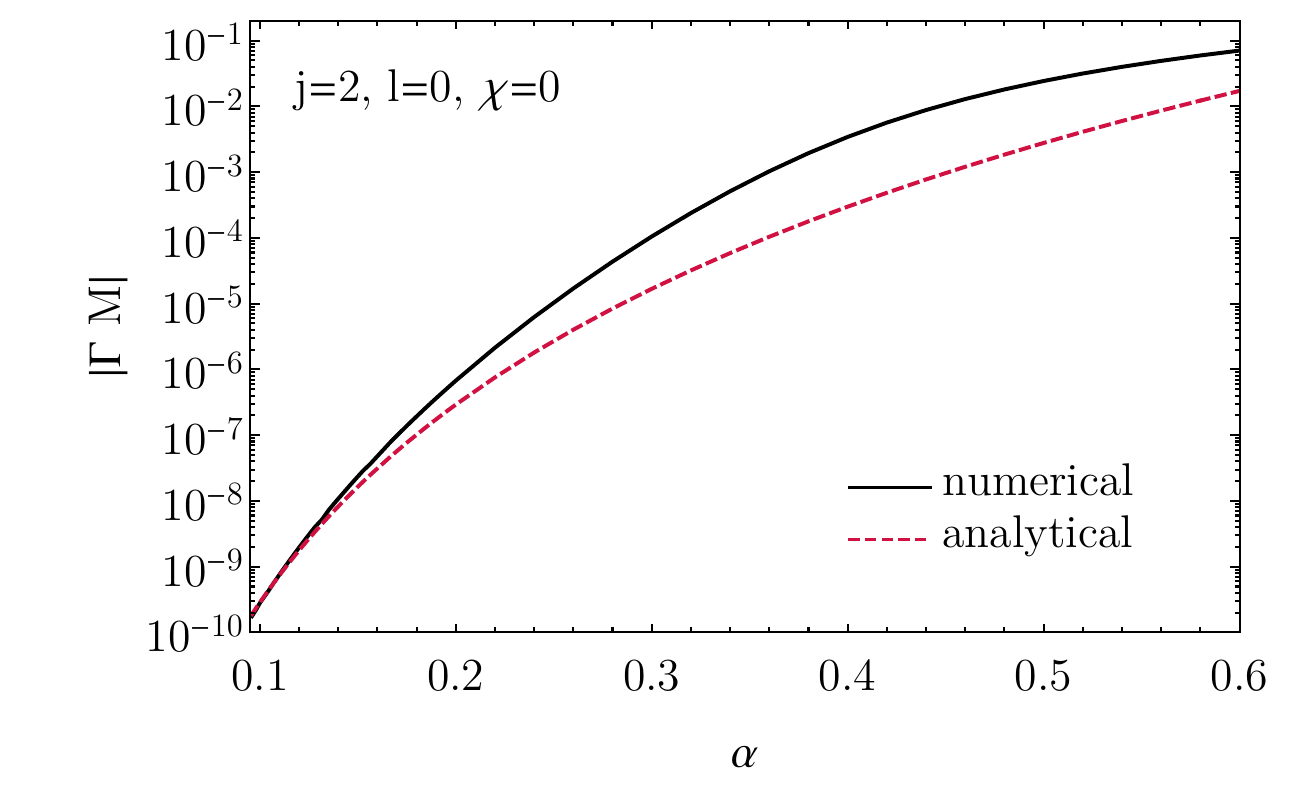}
\includegraphics[width=0.49\textwidth]{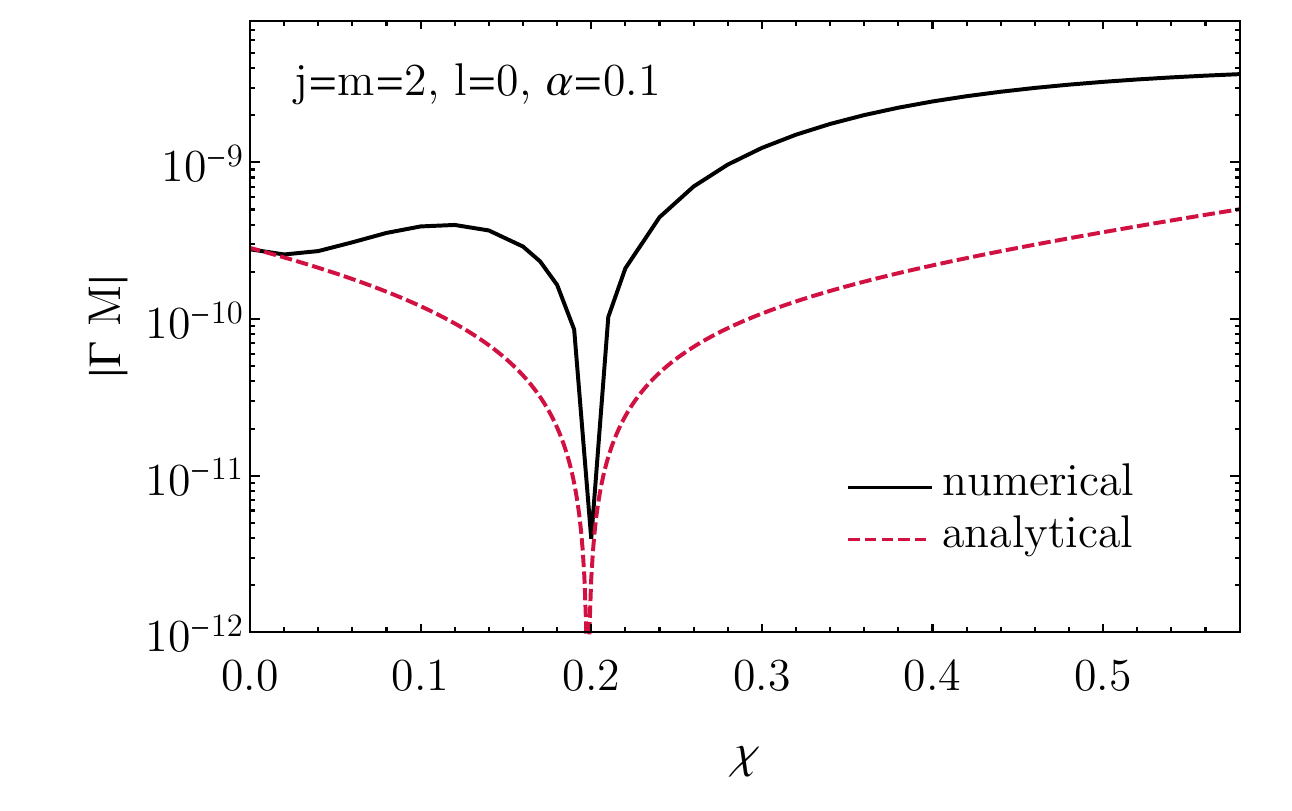}
\caption{Comparison between the analytical decay/growth rates computed in this paper (valid for $\alpha\ll 1$ and any 
spin) and the numerical rates computed in Ref.~\cite{Brito:2013wya} (valid for $\chi\ll 1$ and any value of $\alpha$). 
In the left panel we show the decay rate for $j=2$, $\ell=0$ and zero spin ($\chi=0$), whereas in the right panel we 
show the decay/growth rate as a function of the spin for $\alpha=0.1$ and $j=m=2$, $\ell=0$. As expected, our 
analytical results are in good agreement with the numerical results for $\alpha\ll 1$ and $\chi \ll 1$.
\label{fig:rates}}
\end{figure*}

Finally, to leading order the spin dependence of the decay rate is entirely encoded in that of the absorption 
probability, Eq.~\eqref{Pabs}. Therefore, to compute the leading-$\alpha$ decay rate for the bound state around a Kerr 
BH of any spin ${\chi}=J/M^2$, we can simply
multiply the Schwarzschild decay rate by the ratio 
$\mathbb{P}_{|{\chi}}/\mathbb{P}_{|{\chi}=0}$~\cite{Baryakhtar:2017ngi}. This yields
\begin{equation}\label{eq:j2l0}
    \Gamma_{|{\chi}}^{\ell=0,j=m=2} = - \frac{128}{45} \alpha^9 \mu \frac{\mathbb{P}_{|{\chi}}}{\mathbb{P}_{|{\chi}=0}} 
\simeq \frac{64}{45} {\chi} \alpha^8 \mu\,,
\end{equation}
where, for clarity, only in the last step we have assumed $\chi\ll1$.
In general, we obtain an expression as in the main text, namely
\begin{equation}
 \Gamma = -C_{j\ell}\frac{{\cal P}_{jm}(\chi)}{{\cal P}_{jm}(0)}\alpha^{2(\ell+j)+5}(\omega_R-m\Omega_{\rm 
H})\,, \label{Gamma_app}
\end{equation}
with $C_{20}= 128/45$. In 
Fig.~\eqref{fig:rates} we compare 
the rates given by Eq.~\eqref{eq:j2l0} with the numerical rates of Ref.~\cite{Brito:2013wya}, which are exact only in 
the 
limit $\chi \ll 1$. As expected Eq.~\eqref{Gamma} agrees very well with the numerical rates when $\alpha\ll 1$ and $\chi 
\ll 1$. On the other hand, for large spins we expect that Eq.~\eqref{eq:j2l0} is a better approximation to the growth 
rate than the results of Ref.~\cite{Brito:2013wya}.

For $j=\ell=m=1$ the same procedure yields $\Gamma_{|{\chi}=0} \simeq -\frac{10}{9} 
\alpha^9 \mu$ in the nonspinning case, and therefore 
$\Gamma_{|{\chi}} \simeq \frac{5}{18} {\chi}\alpha^8 \mu$. The full expression is Eq.~\eqref{Gamma} with $C_{11}=10/9$. 
We note that in this case we set $s=1$ in Eq.~\eqref{Pabs}, since the dipole mode was obtained using~\eqref{eq:YE1} 
which transforms as a vector~\cite{Thorne:1980ru}.

Finally, for completeness we have also computed the instability time scale for $j=3$, $\ell=1$ which gives 
Eq.~\eqref{Gamma} with $C_{31}={4}/{4725}$, and for $j=1$, $\ell=2$, which gives $C_{12}=640/19683$.
The latter exactly matches the analytical results of Ref.~\cite{Brito:2013wya} [cf. their 
Eq.~(48)], after noting that the results in~\cite{Brito:2013wya} refer to the decay rate of the field's amplitude, 
whereas our definition refers to the decay rate of the energy density, which is twice the decay rate of the amplitude.

Our main results for the instability time scale are summarized in Table~\ref{tab:inst}.

\begin{table}\label{tab:inst}
 \caption{Coefficients appearing in Eq.~\eqref{Gamma} for the decay rate $\Gamma$ of massive spin-2 modes around a Kerr 
BH. As a reference, in the last column we report the instability time scale $\tau_{\rm inst}=1/\Gamma$ for 
$M=30M_\odot$, $\chi=0.8$, and $\alpha=0.2$.}
 \begin{tabular}{cc|c|c}
  \hline
  \hline
  $j$ 	& 	$\ell$ 	& 	$C_{j\ell}$	& $\tau_{\rm inst}\,[{\rm s}]$ \\
  \hline
  $2$	&	 $0$	&	$128/45$	& $1.0\times10^3$				\\
  $1$	&	 $1$	&	$10/9$		& $2.8\times10^4$				\\
  $3$	&	 $1$	&	$4/4725$	& $3.6\times10^8$				\\
   $1$	&	 $2$	&	$640/19683$	& $2.4\times10^7$				\\
  \hline
  \hline
 \end{tabular}
\end{table}

%

\subsection{GW emission from the condensate}
Let us now discuss the derivation of the GW emission from the condensate.
From the action given in the main text we get that the massive spin-2 condensate sources GWs through the usual equation:
\begin{equation}
	\mathcal{E}_{ab}^{cd} G_{cd} = 8\pi T_{ab}+\mathcal{O}(G^2) 	\label{eq:GW_eom}\,,
\end{equation} 
where $\mathcal{E}_{ab}^{cd} G_{cd}$ is the linearized Einstein's operator and $T_{ab}$ is the stress-energy tensor 
given in Eq.~\eqref{Tmunu} which, when $\alpha \ll 1$, can be computed using the hydrogenic solutions for the unstable 
modes. In this limit the condensate is localized far away from the BH, $r_{\rm Bohr}\gg M$, and GW emission can 
be approximately analyzed in a flat background~\cite{Yoshino:2013ofa}.  

The gravitational radiation sourced by the condensate is best described using the Teukolsky 
formalism~\cite{Teukolsky:1973ha,Yoshino:2013ofa,Brito:2014wla}, which allows us to separate the field 
equations into a system of ordinary differential equations. In this formalism, gravitational radiation 
is described by the Newman-Penrose scalar $\psi_4$ which, in the flat space approximation, can be decomposed as
\begin{equation}
\label{psi4}
   \psi_4(t,r,\Omega)=\sum_{j=0}^{\infty}\sum_{m=-j}^{j} \int^\infty_{-\infty} \frac{R_{jm\omega}(r)}{r^4}~_{-2}Y_{jm}(\Omega)e^{-i\omega t}d\omega \;,
\end{equation}
where ${_s}Y_{jm}(\theta,\phi)$ are the spin-$s$ weighted spherical harmonics. The radial function $R_{jm \omega}(r)$ 
satisfies the inhomogenous Teukolsky equation
\begin{equation}
\begin{aligned}
\label{Teukolskyradialeq}
 &r^2 R_{jm\omega}''-2(r-M)R_{jm\omega}'+[\omega^2r^2-4i\omega (r-3M)\\
 &-(j-1)(j+2)]R_{jm\omega}=T_{jm\omega} \;.
\end{aligned}
\end{equation}

The source term $T_{jm\omega}$ is related to the stress-energy tensor $T_{ab}$ through the tetrad projections 
$T_{ab}n^{a}n^{b}\equiv T_{nn}$, $T_{ab}n^{a}\bar{m}^{b}\equiv T_{n\bar{m}}$ and $T_{ab}\bar{m}^{a}\bar{m}^{b}\equiv 
T_{\bar{m}\bar{m}}$, where we adopted the Kinnersley tetrad, that in Boyer-Lindquist coordinates reads
\begin{align}
&n^{\mu}=\frac{1}{2}\left (1,-1,0,0 \right )\;,\\
&\bar{m}^{\mu}=\frac{1}{\sqrt{2}\,r}\left (0,0,1,-\frac{i}{\sin\vartheta}
\right )\; .
\end{align}
By defining
\begin{equation}\label{source_int}
 _{S}T\equiv \frac{1}{2\pi} \int\, d\Omega\, dt \, {\cal T}_{S}~_{S}\bar{Y}_{jm}e^{i\omega t}\;,   
\end{equation}
where ${\cal T}_S=T_{nn}$, $T_{n\bar{m}}$ and $T_{\bar{m}\bar{m}}$ for $S=0,-1,-2$ respectively, the source reads 
\begin{eqnarray}
\frac{T_{jm\omega}}{2\pi}&=&2\left[(j-1)j(j+1)(j+2)\right]^{1/2}r^4~_{0}T \notag\\
&+&2\left[2(j-1)(j+2)\right]^{1/2}r^2  \mathcal{L}\left(r^3 ~_{-1}T\right)\notag\\
&+&r \mathcal{L}\left[r^4 \mathcal{L}\left(r~_{-2}T\right)\right]\,,
\end{eqnarray}
where $\mathcal{L}\equiv \partial_r+i\omega$.

Once the source term is known, the radial equation ~\eqref{Teukolskyradialeq} can be solved using the Green's function, which in the small-frequency limit can be used to find analytical solutions~\cite{Poisson:1993vp,Ficarra:2018rfu,Brito:2014wla} as we now describe.
To construct the Green's function we consider two linearly independent solutions of the homogeneous equation. A physically motivated choice is to consider the solution $R^{\infty}$ which describes outgoing waves at infinity and $R^H$ which describes ingoing waves at the event horizon: 
\begin{equation}
R^{H} \to   \begin{cases}
   r^4 e^{-i\omega r} \hspace{4cm} r \to 0\,,\\r^3B_{\rm  out} e^{i\omega r} + r^{-1} B_{\rm  in} e^{-i\omega r} \hspace{1.2cm} r 
\to \infty \,,
   \end{cases}
\end{equation}
\begin{equation}
R^{\infty} \to   \begin{cases}
   A_{\rm out} e^{i\omega r} + r^4 A_{\rm  in} e^{-i\omega r} \hspace{1.8cm} r \to 0\,,\\r^3 e^{i\omega r}  \hspace{4.3cm} r \to 
\infty \,,
   \end{cases}
\end{equation}
where we note that --~owing to the flat space approximation~-- the event horizon lies at $r\to 0$.

Imposing ingoing boundary conditions at the horizon and outgoing boundary conditions at infinity, we find that the 
solution
of Eq.~\eqref{Teukolskyradialeq} is given by
\begin{equation}\label{eq:full_sol}
R_{jm\omega}(r)= \frac{R^{\infty}}{W}\int_{0}^{r} dr' \frac{R^H T_{jm\omega}}{r'^4} + \frac{R^{H}}{W}\int_{r}^{\infty} 
dr' \frac{R^{\infty} T_{jm\omega}}{r'^4} \;,
\end{equation}
where the Wronskian $W=2i \omega B_{\rm in}$ is a constant. The constant $B_{\rm in}$ can be found through the asymptotic solution of the homogeneous Teukolsky equation~\eqref{Teukolskyradialeq} and reads
\begin{eqnarray}
B_{\rm in}=-\frac{C_1}{8\omega^2}(j-1)j(j+1)(j+2)e^{i(j+1)\frac{\pi}{2}}\;,
\end{eqnarray}
where $C_1$ is an arbitrary constant that we can set to unity
without loss of generality. The solution $R^H$ can be found through $R^H=r^2\mathcal{L}\left(\mathcal{L}r 
\psi^H\right)$,
where $\psi^H$ is the Regge-Wheeler function that, at small frequencies, reads
\begin{eqnarray}
\psi^H\sim \omega r j_j(\omega r)\,.
\end{eqnarray}

Since we are interested in gravitational radiation measured at very large distances from the system, we are interested in the solution~\eqref{eq:full_sol} when $r\to\infty$,  which reads
\begin{eqnarray}
\label{Ztilde}
R_{jm\omega}(r\to \infty)&\sim &\frac{R^{\infty}(r\to \infty)}{2i\omega B_{\rm in}}\int_{0}^{\infty} dr' \frac{R^H(r') T_{jm\omega}(r')}{r'^4} \nonumber\\
&\equiv& \tilde{Z}_{jm\omega}^{\infty}r^3 e^{i\omega r}\;.
\end{eqnarray}

From Eq.~\eqref{source_int} ones finds that, for a condensate dominated by a single mode with frequency $\omega_R$, the 
frequency spectrum of the source $T_{jm\omega}$ is discrete with frequencies $\tilde{\omega}_{1}=+ 2\omega_{R}$ and 
$\tilde{\omega}_{2}=- 2\omega_{R}$. Therefore $\tilde{Z}_{jm\omega}^{\infty}$ in Eq.~\eqref{Ztilde} can be written as
\begin{equation}
  \tilde{Z}_{jm\omega}^{\infty}=  \sum_{n=1}^2 \delta(\omega-\tilde{\omega}_{n}) Z_{jm\tilde{\omega}_{n}}^{\infty} \;.
\end{equation}
Replacing the above equation in Eq.~\eqref{psi4} we obtain, at $r\to\infty$,
\begin{equation}
\label{psi4infty}
    \psi_{4} = \frac{1}{r}
     \sum_{j=0}^{\infty}\sum_{m=-j}^{j} \sum_{n=1}^2 
Z_{jm\tilde{\omega}_{n}}^{\infty}~_{-2}Y_{jm}\,e^{i\tilde{\omega}_{n}(r- t)} \;,
\end{equation}
which is related to the two independent GW polarizations $h_{+}$ and $h_{\times}$ by
\begin{equation}
    \psi_4=\frac{1}{2} \left( \ddot{h}_{+} - i \ddot{h}_{\times}  \right) \;.
\end{equation}
Using Eq.~\eqref{psi4infty} and integrating twice with respect to the time, we obtain
\begin{equation}\label{waveform}
    h_{+} -i h_{\times} = \frac{2}{r}
     \sum_{j=0}^{\infty}\sum_{m=-j}^{j} \sum_{n=1}^2 
\frac{Z_{jm\tilde{\omega}_{n}}^{\infty}}{\tilde{\omega}_{n}^2}~_{-2}Y_{jm}\,e^{i\tilde{\omega}_{n}(r- t)} \;.
\end{equation}

The energy flux carried by these waves at radial infinity is given by 
\begin{equation}
    \frac{dE_{\rm GW}}{dtd\Omega} =\sum_{n=1}^2 \frac{r^2}{4\pi\tilde{\omega}_n^2} |\psi_4|^2 \equiv \frac{r^2}{16\pi} \left( 
\dot{h}^2_{+} + \dot{h}^2_{\times}  \right) \;.
\end{equation}
Finally, using~\eqref{waveform}, we get the energy flux
\begin{equation}
     \dot{E}_{\rm GW}= \sum_{jmn} \frac{1}{4\pi \tilde{\omega}_{n}^2} |Z_{lm\tilde{\omega}_{n}}^{\infty}|^2 \;,
\end{equation}
where the dot represents the time derivative.

Considering a condensate dominated by a single massive spin-2 mode with total angular momentum $j$ and azimuthal 
number $m$, Eq.~\eqref{source_int} implies that the gravitational radiation is emitted in
angular modes with $j_{\rm GW}=2j$ and $m_{\rm GW}=\pm 2m$, where the plus and minus sign correspond to the frequency 
$\tilde{\omega}_{1}=+ 2\omega_{R}$ and $\tilde{\omega}_{2}=- 2\omega_{R}$, respectively. 

Employing the procedure outlined above we find that, at the leading order in $\alpha$, the GW flux for the most 
unstable modes is given by
\begin{align}
 \dot{E}_{\rm GW}^{\ell=0,j=m=2}&= \frac{1048576+38025\pi^2}{28350} \left(\frac{M_{c}}{M}\right)^2 \alpha^{10}\;,\\
 \dot{E}_{\rm GW}^{j=\ell=m=1} &= \frac{1530169+1687401\pi^2}{46656000} \left(\frac{M_{c}}{M}\right)^2 \alpha^{14}\;,
\end{align}
where $M_c$ is the total mass of the condensate, which can be related to the amplitude of the massive spin-2 field 
through Eq.~\eqref{energy_condensate}. 

To estimate $M_c$ we follow Ref.~\cite{Brito:2017zvb}. Since in the small $\alpha$ limit the instability time scale and 
GW emission time scale satisfy $\tau_{\rm GW}\gg \tau_{\rm inst}$, the process occurs in two stages: first the 
condensate grows in a time scale $\tau_{\rm inst}$ with negligible emission of GWs until the system saturates at 
$\omega_R=m\Omega_{\rm H}$; then, on a longer time scale, the system slowly dissipates through the emission of GWs. This 
picture was shown to describe very well the evolution of the system even beyond the small-$\alpha$ 
limit~\cite{East:2017ovw}. Therefore, we can safely neglect GW emission to compute $M_c$ at the end of the superradiant 
phase. Conservation of energy and angular momentum imply that the BH mass $M_f$ and spin $J_f$ at the end of the 
superradiant growth phase are related to the initial BH mass $M_i$ and spin $J_i$ through the relation
\begin{equation}
J_f=J_i -\frac{m}{\omega_R}\left(M_i-M_f\right)\,.
\end{equation}
Since the superradiant growth stops when $\omega_R/m=\Omega_{\rm H}(M_f,J_f)$ we can combine both equations to get~\cite{Tsukada:2018mbp}
\begin{equation}
M_f=\frac{m^3-\sqrt{m^6-16m^2\omega_R^2\left(m M_i-\omega_R J_i \right)^2}}{8\omega_R^2\left(m M_i-\omega_R J_i\right)}\,,
\end{equation}
where we remind that $\omega_R\sim \mu$ in the $\alpha\ll 1$ limit.
For a given BH with mass $M_i$ and spin $J_i$ the maximum mass of the condensate can then be computed using $M_c= M_i-M_f$.
%

\section{Self-interactions}
\label{sec:selfint}
The nonlinear interaction of two spin-2 fields is described by the unique general action of bimetric 
theories~\cite{Hassan:2011hr,Hassan:2011zd}, which also depends on the relative coupling 
$\epsilon=M_P^{(2)}/M_P$ of the second metric to matter, where $M_P$ and $M_P^{(2)}$ are the effective Planck masses 
associated to the first and second metric, respectively. 
The relative coupling $\epsilon$ is essentially unconstrained 
for ultralight spin-2 fields~\cite{Babichev:2016bxi}. Since we are interested in dark-matter fields (i.e., in fields 
that interact mostly gravitationally with the Standard Model particles), we require $\epsilon\ll1$.
The spectrum of vacuum solutions of bimetric theory is very rich 
and larger than the Kerr family~\cite{Babichev:2015xha}. 
However, in the $\epsilon\ll1$ limit the general-relativistic solutions 
are all recovered either approximately~\cite{Babichev:2016bxi} (with small ${\cal O}(\epsilon)$ corrections) or exactly 
(when the two background metrics are proportional to each other), so our assumption of a Kerr 
background metric is well justified.

In the $\epsilon\ll1$ limit, the terms in the cubic Lagrangian ${\cal L}_{\rm cubic}$ of the action given in the main text are cubic in $H_{ab}$ and can be schematically written in two forms:~\cite{Babichev:2016bxi}
\begin{equation}
 \frac{m_b^2}{\epsilon}{\cal O}(H^3)\,,\qquad \frac{1}{\epsilon}{\cal O}(H (\nabla H)^2)\,. \label{orders}
\end{equation}
For the hydrogenic unstable modes, $\nabla H\sim H/r_{\rm Bohr}$ so that in the small-coupling limit the second terms 
in Eq.~\eqref{orders} are smaller than the first ones. By comparing the leading cubic terms with the mass terms for 
$H_{ab}$ at the level of the action~\cite{Ikeda:2019fvj}, we can estimate that nonlinearities become important when
\begin{equation}
 |H_{ab}|\gtrsim \epsilon M_P = M_P^{(2)}\,,
\end{equation}
i.e. when the massive spin-2 field is of the order of the effective Planck mass associated to the second metric. Since 
$\epsilon$ is essentially unconstrainted, the nonlinear scale is parametrically smaller than $M_P$ but can nonetheless 
be very high. As long as $|H_{ab}|\ll \epsilon M_P$ our analysis 
(which neglects nonlinearities in $H_{ab}$) is robust. In the opposite regime, interesting novel effects such as 
``bosenovas''~\cite{Yoshino:2012kn,Yoshino:2015nsa} from massive spin-2 fields and formation of geons~\cite{Aoki:2017ixz} with self-interactions can occur.

\end{document}